\documentclass[a4paper,twoside,twocolumn,english,aps,floatfix,superscriptaddress,showpacs]{revtex4}
\usepackage{times}
\usepackage[T1]{fontenc}
\usepackage[latin1]{inputenc}
\usepackage{amsmath}
\usepackage{graphicx}
\usepackage{amssymb}

\makeatletter

\newcommand{\noun}[1]{\textsc{#1}}

\usepackage{babel}
\makeatother
\begin{document}

\title{Excitations of Few-Boson Systems in 1-D Harmonic and Double Wells}

\author{Sascha Zöllner}

\email{sascha.zoellner@pci.uni-heidelberg.de}

\affiliation{Theoretische Chemie, Institut f\"{u}r Physikalische Chemie, Universit\"{a}t
Heidelberg, INF 229, 69120 Heidelberg, Germany}

\author{Hans-Dieter Meyer}

\email{hans-dieter.meyer@pci.uni-heidelberg.de}

\affiliation{Theoretische Chemie, Institut f\"{u}r Physikalische Chemie, Universit\"{a}t
Heidelberg, INF 229, 69120 Heidelberg, Germany}

\author{Peter Schmelcher}

\email{peter.schmelcher@pci.uni-heidelberg.de}

\affiliation{Theoretische Chemie, Institut f\"{u}r Physikalische Chemie, Universit\"{a}t
Heidelberg, INF 229, 69120 Heidelberg, Germany}

\affiliation{Physikalisches Institut, Universit\"{a}t Heidelberg, Philosophenweg
12, 69120 Heidelberg, Germany}

\begin{abstract}
We examine the lowest excitations of one-dimensional few-boson systems
trapped in double wells of variable barrier height. Based on a numerically
exact multi-configurational method, we follow the whole pathway from
the non-interacting to the fermionization limit. It is shown how,
in a purely harmonic trap, the initially equidistant, degenerate levels
are split up due to interactions, but merge again for strong enough
coupling. In a double well, the low-lying spectrum is largely rearranged
in the course of fermionization, exhibiting level adhesion and (anti-)crossings.
The evolution of the underlying states is explained in analogy to
the ground-state behavior. Our discussion is complemented by illuminating
the crossover from a single to a double well.
\end{abstract}

\date{\today}


\maketitle

\section{Introduction}

The realization of Bose-Einstein condensates has made ultracold atoms
an ideal tool for probing and understanding paradigm quantum phenomena
\cite{pitaevskii,dalfovo99,pethick,leggett01}. One such example is
the quasi-one-dimensional Bose gas, where the transverse degrees of
freedom are frozen out such that an effective one-dimensional description
becomes possible. As it turns out, in such a system one can tune the
effective atom-atom interaction strength at will by merely changing
the transverse confinement length \cite{Olshanii1998a}. This allows
us to explore the limit of strong correlations and, in this way, to
test the physics beyond the mean-field description applicable to large
and \emph{weakly} interacting systems.

The limit of \emph{infinite} repulsion---so-called hard-core bosons---is
in turn known exactly in one dimension, as it is isomorphic to that
of \emph{free} fermions (up to permutation symmetry \cite{girardeau60}).
This makes it appealing to think of the exclusion principle as mimicking,
as it were, the hard-core repulsion, which is why this limit is termed
\emph{fermionization}. Its theoretical description has thus attracted
a great deal of attention \cite{vaidya79,minguzzi02,girardeau01,papenbrock03},
stimulated by its recent experimental verification \cite{kinoshita04,paredes04}.

On the other hand, the question how exactly these two very different
borderline cases connect has attracted less attention. The major reason
for this is that it is notoriously hard to include the effects of
strong correlations from first principles. This can only be done for
rather few particles, say $N\sim10$. As it happens, this is not only
the regime of experimental accessibility \cite{petrov00}; it is also
very interesting because the signatures of two-body interactions are
still pronounced, thus facilitating the understanding of larger systems.
An analytic solution is known for homogeneous systems with periodic
\cite{lieb03,sakmann05} and fixed boundaries \cite{hao06}, but also
for the elementary case of two atoms in a harmonic trap \cite{Busch98},
which already captures some key features of the evolution from the
weakly to the strongly interacting limit. In general, though, the
solution of trapped interacting bosons requires numerical approaches.
Most investigations have so far focused primarily on the ground state
\cite{blume02,alon05,deuretzbacher06,streltsov06,zoellner06a,zoellner06b},
or on regimes where correlations are weak enough to be passably represented
by few single-particle orbitals \cite{masiello05,masiello06,cederbaum04}. 

The goal of this paper now is to extend the systematic study of the
fermionization transition in harmonic and double-well traps to the
lowest excitations of finite boson systems. Understanding the low-lying
spectral properties is not only an interesting problem in its own
right, given the richness of the pathway to fermionization for the
ground state. It is also an essential contribution to the control
of few-body systems (as is desirable for quantum information processing)
and to explaining their dynamics, where the double well is a prototype
model for fundamental effects such as tunneling or interference \cite{andrews97,shin04,anker05}.
As in our previous works \cite{zoellner06a,zoellner06b}, we draw
on the Multi-Configuration Time-Dependent Hartree method \cite{mey03:251,mey98:3011,bec00:1}.
As its single-particle basis is variationally optimized for each state,
it allows us to study the excited states of few bosons in a numerically
exact way.

Our paper is organized as follows. Section~\ref{sec:theory} introduces
the model and gives an overview of some key concepts. In Sec.~\ref{sec:method},
we give a brief introduction to the computational method and how it
can be applied to the study of excitations. The subsequent section
finally is devoted to our results for the low-lying spectrum (Sec.~\ref{sub:spectrum})
and the underlying excited states (Sec.~\ref{sub:states}). Our discussion
is rounded off by illuminating the crossover from a single well to
a double-well trap in Sec.~\ref{sub:barrier-crossover}.

\section{Theoretical background \label{sec:theory}}

\subsection{Model}

 In this work we investigate a system of \emph{few} interacting bosons
($N=2,\dots,5$) in an external trap. These particles, representing
atoms, are taken to be one-dimensional (1D). More precisely, after
integrating out the transverse degrees of freedom and upon introducing
\emph{dimensionless} variables we arrive at the model Hamiltonian
(see \cite{zoellner06a} for details) \[
H=\sum_{i}h_{i}+\sum_{i<j}V(x_{i}-x_{j}),\]
where $h=\frac{1}{2}p^{2}+U(x)$ is the one-particle Hamiltonian with
a trapping potential $U$, while $V$ is the effective two-particle
interaction potential \cite{Olshanii1998a}\[
V(x)=g\delta_{\sigma}(x),\textrm{ with }g=\frac{2a_{0}}{a_{\perp}^{2}}\left(1-\left|\zeta\left({\scriptstyle \frac{1}{2}}\right)\right|\frac{a_{0}}{a_{\perp}}\right)^{-1}\negthickspace.\]
Here an s-wave scattering length $a_{0}$ and a harmonic transverse
trap with oscillator length $a_{\perp}$ were assumed. The well-known
numerical difficulties due to the spurious short-range behavior of
the standard delta-function potential $\delta(x)$ are alleviated
by mollifying it with the normalized Gaussian\[
\delta_{\sigma}(x)=\frac{1}{\sqrt{2\pi}\sigma}e^{-x^{2}/2\sigma^{2}},\]
which tends to $\delta(x)$ as $\sigma\to0$ in the distribution sense.
We choose a fixed value $\sigma=0.05$ as a trade-off between smoothness
and a range that is much shorter than the length scale of the trap,
$a_{\parallel}=1$.

\subsection{Correlations and fermionization: key aspects\label{sub:Fragmentation}}

Our approach equips us with the full solution of the system---here,
the excited-state wave functions, which are fairly complex entities.
Visualizing and in this way relating them to the physical picture,
it is useful to consider reduced densities, or correlation functions.
As is well-known, the knowledge of some wave function $\Psi$ is equivalent
to that of the density matrix $\rho_{N}=|\Psi\rangle\langle\Psi|$.
To the extent that we study at most two-body correlations, it already
suffices to consider the reduced two-particle density operator\begin{equation}
\rho_{2}=\mathrm{tr}_{3..N}|\Psi\rangle\langle\Psi|,\label{eq:rho2}\end{equation}
whose diagonal kernel $\rho_{2}(x_{1},x_{2})$ gives the probability
density for finding one particle located at $x_{1}$ and any second
one at $x_{2}$. For any one-particle operator, of course, it would
be enough to know the one-particle density matrix $\rho_{1}=\mathrm{tr}_{2}\rho_{2}$,
so that the exact energy may be written as\[
E=N\,\mathrm{tr}(\rho_{1}h)+\frac{N(N-1)}{2}\mathrm{tr}(\rho_{2}V).\]
 The one-particle density matrix can be characterized by its spectral
decomposition \begin{equation}
\rho_{1}\equiv\sum_{a}n_{a}|\phi_{a}\rangle\langle\phi_{a}|,\label{eq:rho1}\end{equation}
 where $n_{a}\in[0,1]$ is said to be the population of the \textit{natural
orbital} $\phi_{a}$. If the system is in a number state $|\boldsymbol{n}\rangle\equiv|n'_{0},n'_{1},\dots\rangle$
based on the one-particle basis $\{\phi_{a}\}$, then $n_{a}\equiv n'_{a}/N$
in (\ref{eq:rho1}); but it also extends that concept to non-integer
values of $n_{a}'$.

Remarkably enough, not only the non-interacting limit is well known,
but also the complementary case of infinitely strong correlations,
$g\to\infty$. It is commonly referred to as the \emph{Tonks-Girardeau}
limit of 1D hard-core bosons, or also as their \emph{fermionization}.
This lingo finds its justification in the \emph{Bose-Fermi map} \cite{girardeau60,yukalov05}
that establishes an isomorphy between the exact \emph{bosonic} wave
function $\Psi_{\infty}^{+}$ and that of a (spin-polarized) non-interacting
\emph{fermionic} solution $\Psi_{0}^{-}$, \begin{equation}
\Psi_{\infty}^{+}=A\Psi_{0}^{-},\label{eq:BF}\end{equation}
where $A=\prod_{i<j}\mathrm{sgn}(x_{i}-x_{j})$. The mapping holds
not only for the ground state, but also for excited and time-dependent
states. Since $A^{2}=1$, their (diagonal) densities as well as their
energy $E$ will coincide with those of the corresponding free fermionic
states. That makes it tempting to think of the exclusion principle
as mimicking the interaction ($g\to\infty$), as is nicely illustrated
on the ground state of $N$ hard-core bosons in a harmonic trap \cite{girardeau01}\[
\Psi_{\infty}^{+}(Q)\propto e^{-|Q|^{2}/2}\negthickspace\prod_{1\le i<j\le N}\negthickspace|x_{i}-x_{j}|,\]
 where $Q=(x_{1},\dots,x_{N})^{\top}$.

\section{Computational method\label{sec:method}}

Our goal is to investigate the lowest excited states of the system
introduced in Sec.~\ref{sec:theory} for all relevant interaction
strengths in a numerically \emph{exact, i.e.,} controllable fashion.
This is a highly challenging and time-consuming task, and only few
such studies on ultracold atoms exist even for model systems (see,
e.g., \cite{streltsov06,masiello05}). Our approach relies on the
Multi-Configuration Time-Dependent Hartree \noun{(mctdh)} method
\cite{bec00:1}, primarily a wave-packet dynamics tool known for its
outstanding efficiency in high-dimensional applications. To be self-contained,
we will provide a concise introduction to this method and how it can
be adapted to our purposes.

\subsection{Principal idea}

The underlying idea of MCTDH is to solve the time-dependent Schrödinger
equation\[
\begin{array}{c}
i\dot{\Psi}=H\Psi,\quad\left.\Psi\right|_{t=0}=\Psi^{(0)}\end{array}\]
 as an initial-value problem by expansion in terms of direct (or Hartree)
products $\Phi_{J}\equiv\varphi_{j_{1}}^{(1)}\otimes\cdots\otimes\varphi_{j_{N}}^{(N)}$:\begin{equation}
\Psi(Q,t)=\sum_{J}A_{J}(t)\Phi_{J}(Q,t).\label{eq:mctdh-ansatz}\end{equation}
The (unknown) \emph{single-particle functions} $\varphi_{j_{\kappa}}^{(\kappa)}$
($j_{\kappa}=1,\dots,n_{\kappa}$) are in turn represented in a fixed
\emph{primitive} basis implemented on a grid. For indistinguishable
particles as in our case, the sets of single-particle functions for
each degree of freedom $\kappa=1,\dots,N$ are of course identical
(i.e., we have $\varphi_{j_{\kappa}}$, with $j_{\kappa}\le n$).

Note that in the above expansion, not only the coefficients $A_{J}$
are time-dependent, but so are the Hartree products $\Phi_{J}$. Using
the Dirac-Frenkel variational principle, one can derive equations
of motion for both $A_{J},\varphi_{j}$ \cite{bec00:1}. Integrating
this differential-equation system allows one to obtain the time evolution
of the system via (\ref{eq:mctdh-ansatz}). Let us emphasize that
the conceptual complication above offers an enormous advantage: the
basis $\{\Phi_{J}(\cdot,t)\}$ is variationally optimal at each time
$t$. Thus it can be kept fairly small, rendering the procedure very
efficient.

It goes without saying that the basis vectors $\Phi_{J}$ are not
permutation symmetric, as would be an obvious demand when dealing
with bosons. This is not a conceptual problem, though, because the
symmetry may as well be enforced on $\Psi$ by symmetrizing the coefficients
$A_{J}$.

\subsection{Application of the method}

The \noun{mctdh} approach \cite{mctdh:package}, which we use, incorporates
a significant extension to the basic concept outlined so far. The
so-called \emph{relaxation method} \cite{kos86:223} provides a way
to not only \emph{propagate} a wave packet, but also to obtain the
lowest \emph{eigenstates} of the system, $\Psi_{m}$. The key idea
is to propagate some wave function $\Psi^{(0)}$ by the non-unitary
$e^{-H\tau}$ (\emph{propagation in imaginary time}.) As $\tau\to\infty$,
this exponentially damps out any contribution but that stemming from
the true ground state like $e^{-E_{m}\tau}$. In practice, one relies
on a more sophisticated scheme termed \emph{improved relaxation} \cite{mey03:251,meyer06},
which is much more viable especially for excitations. Here $\langle\Psi|H|\Psi\rangle$
is minimized with respect to both the coefficients $A_{J}$ and the
orbitals $\varphi_{j}$. This leads to (i) a self-consistent eigenvalue
problem for $(\langle\Phi_{J}|H|\Phi_{K}\rangle)$, which yields $A_{J}$
as {}`eigenvectors' , and (ii) equations of motion for the orbitals
$\varphi_{j}$, but based on certain mean-field Hamiltonians. These
are solved iteratively by first diagonalizing for $A_{J}$ with \emph{fixed}
orbitals and then {}`optimizing' $\varphi_{j}$ by propagating them
in imaginary time over a short period. That cycle will then be repeated.

Whereas the convergence to the ground state is practically bulletproof,
matters are known to get trickier for excited states (see \cite{meyer06}).
This should come as no surprise, granted that one cannot just seek
the energetically lowest state possible but should remain orthogonal
to any neighboring vectors $\Psi_{m}$. (That is why, at bottom, the
convergence turns out to be highly sensitive to the basis size---that
is, to $n$---even for small correlations: The lower states simply
must be represented accurately enough.) For practical purposes, the
most solid procedure has proven to be the following. In the non-interacting
case, we construct the eigenstates as number states $|\boldsymbol{n}\rangle\equiv|n'_{0},n'_{1},\dots\rangle$
in the single-particle basis $\{\phi_{a}\}$. Starting from a given
$|\boldsymbol{n}\rangle$, the eigenstate $\Psi_{m}$ for $g>0$ is
found by an improved relaxation while sieving out the eigenvector
\emph{closest} to its initial state $|\boldsymbol{n}\rangle$. The
resulting eigenstate will then in turn serve as a starting point for
an even larger $g$, and so on.

As it stands, the effort of this method scales exponentially with
the number of degrees of freedom, $n^{N}$. This restricts our analysis
in the current setup to about $N=O(10)$, depending on how decisive
correlation effects are.  Since the computation of excited states
requires that the neighboring states be sufficiently well represented,
the basis must in fact be rather large even for weak correlations. 

As an illustration, we consider systems with $N\sim5$ and need $n\sim15$
orbitals. By contrast, the dependence on the primitive basis, and
thus on the grid points, is not as severe. In our case, the grid spacing
should of course be small enough to sample the interaction potential.
We consider a discrete variable representation with $95$ to $125$
grid points per degree of freedom.

\section{Lowest excitations \label{sec:excited}}

As in Refs.~\cite{zoellner06a,zoellner06b}, we consider bosons in
a double-well trap modeled by\[
U(x)=\frac{1}{2}x^{2}+h\delta_{w}(x),\]
expressed in terms of the harmonic-oscillator length $a_{\parallel}$.
This potential is a superposition of a \emph{}harmonic oscillator
(HO), which it equals asymptotically, and a central barrier which
splits the trap into two fragments. The barrier is shaped as a normalized
Gaussian $\delta_{w}$ of width $w$ and barrier strength $h$. If
$w\to0$, the effect of the barrier reduces to that of a mere boundary
condition (since $\delta_{w}\to\delta$), and the corresponding \emph{one-particle}
problem can be solved analytically \cite{Busch03}. Although this
soluble borderline case presents a neat toy model, the exact width
$w$ does not play a decisive role and is set to $w=0.5$.

In Refs.~\cite{zoellner06a,zoellner06b}, we have studied the ground-state
evolution from the weakly correlated regime to fermionization, with
an eye toward the fascinating interplay between inter-atomic and external
forces as the barrier height $h$ was varied. We now seek to extend
that investigation to the lowest excitations. In Sec.~\ref{sub:spectrum}
we will look into the low-lying spectrum $\sigma(H)=\{ E_{m}\}$,
whose corresponding eigenstates $\Psi_{m}$ will be analyzed in detail
(Sec.~\ref{sub:states}). As the spectral properties in the cases
of a single and a double well will turn out to be quite different,
the question as to how they connect naturally arises. That crossover
will be the subject of \ref{sub:barrier-crossover}.

\subsection{Spectrum \label{sub:spectrum}}

In this section, we study the evolution of the lowest energies $E_{m}(g)$
as $g$ passes from the non-interacting to the fermionization limit.
Figures~\ref{cap:energy_h0},\ref{cap:energy_h5} convey an impression
of this transition for $N=3,4,5$ bosons in a harmonic trap ($h=0$)
and in a double well ($h=5$), respectively. Before dwelling on the
details, let us first capture some universal features of the spectra. 

In the uncorrelated limit, $g\to0$, the energies are simply given
by distributing the atoms over the single-particle levels $\epsilon_{a}$,
starting from $n'_{0}=N$ (the Bose {}`condensate'):\begin{equation}
E=N\,\mathrm{tr}\left(\rho_{1}h\right)=\sum_{a}n_{a}'\epsilon_{a}.\label{eq:spe}\end{equation}
In particular, $E_{0}=N\epsilon_{0}$; hence the {}`chemical potential'
$\mu_{N}\equiv E_{0}^{(N+1)}-E_{0}^{(N)}=\epsilon_{0}$, as usual.
Note that Eq.~(\ref{eq:spe}) implies degeneracy if two single-particle
energies are commensurate, i.e., $\sum_{a}(n'_{a}-\tilde{n}_{a}')\epsilon_{a}=0$
for two $\boldsymbol{n}\neq\tilde{\boldsymbol{n}}$. 

In the Tonks-Girardeau limit, on the other hand, the spectrum becomes
that of a free fermionic system (even though, of course, the system
is really still bosonic and has an all but negligible share of interaction
energy). Thus one can find some (auxiliary) $\boldsymbol{n}$ with
$n_{a}'\in\left\{ 0,1\right\} $ such that \begin{equation}
\lim_{g\to\infty}E(g)=\sum_{a}n_{a}'\epsilon_{a}.\label{eq:tge}\end{equation}
In the ground state, the particles can therefore be thought of as
filling the energy ladder up to the Fermi edge, $\epsilon_{a}<\epsilon_{N}=\mu_{N}$.
For a harmonic confinement, the chemical potential will thus be $\propto N$,
so $E^{(N)}=O(N^{2})$.

It should be pointed out that, in the spirit of the Bose-Fermi map
(\ref{eq:BF}), the borderline cases of no and infinite repulsion
may be perceived as one and the same (non-interacting) system, their
sole difference being the {}`exchange symmetry' emulating the effect
of interactions. Therefore the same type of energy spacings and (quasi-)degeneracies
should appear at both ends of the spectrum.

\subsubsection*{Harmonic trap ($h=0$)}

\begin{figure}
\includegraphics[%
  width=7cm,
  keepaspectratio]{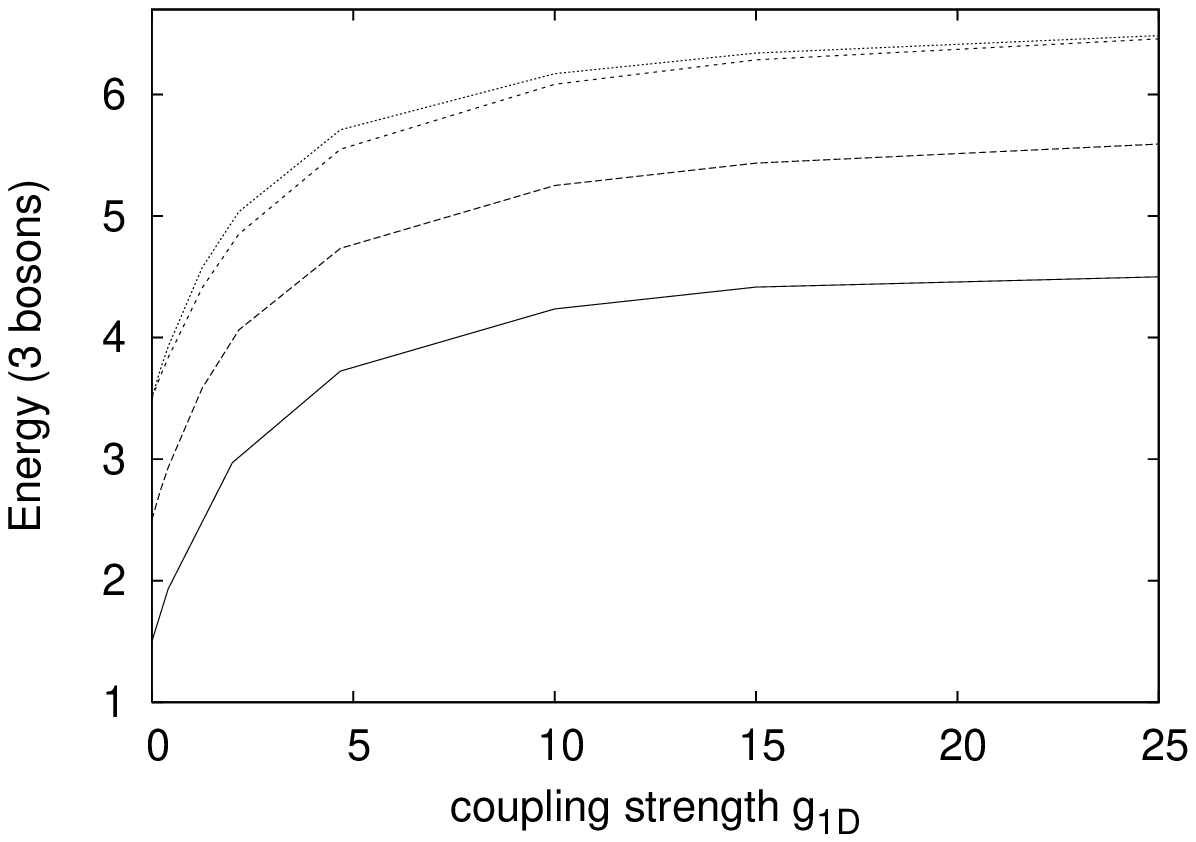}

\includegraphics[%
  width=7cm,
  keepaspectratio]{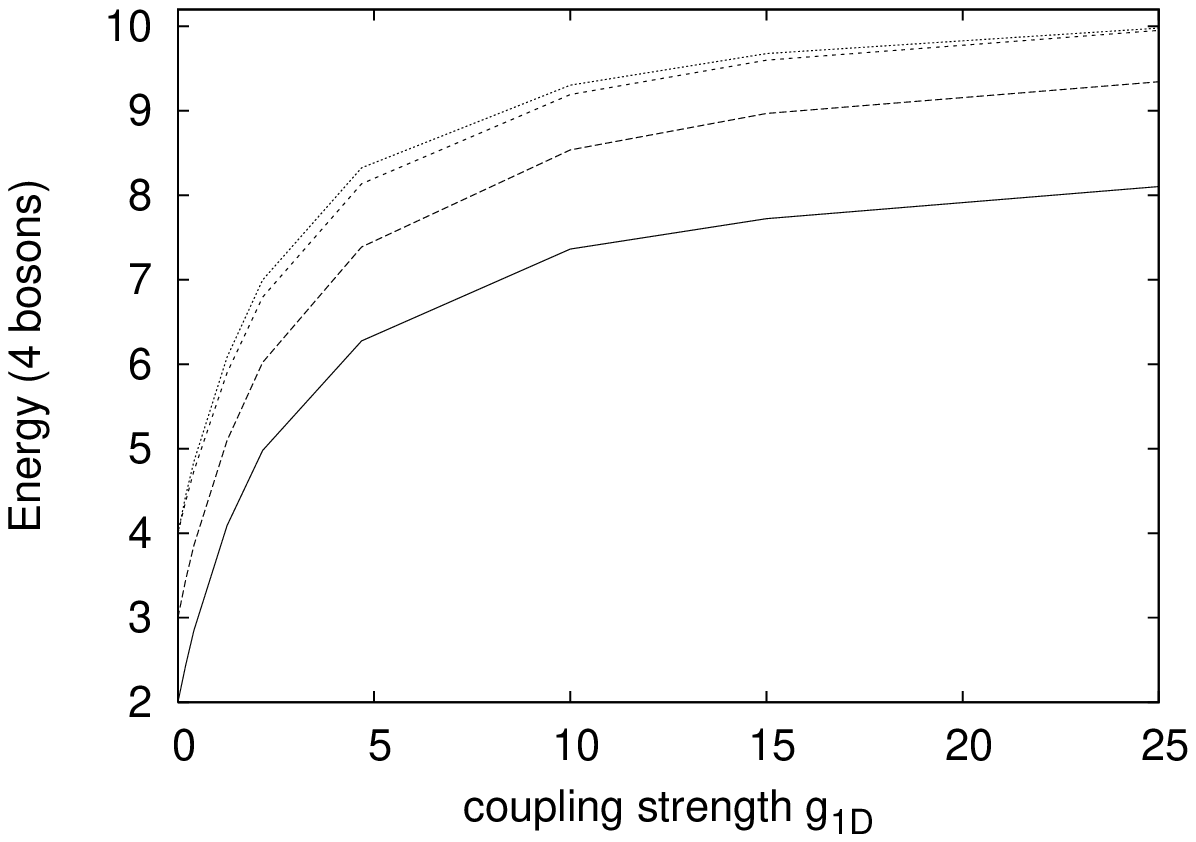}

\includegraphics[%
  width=7cm,
  keepaspectratio]{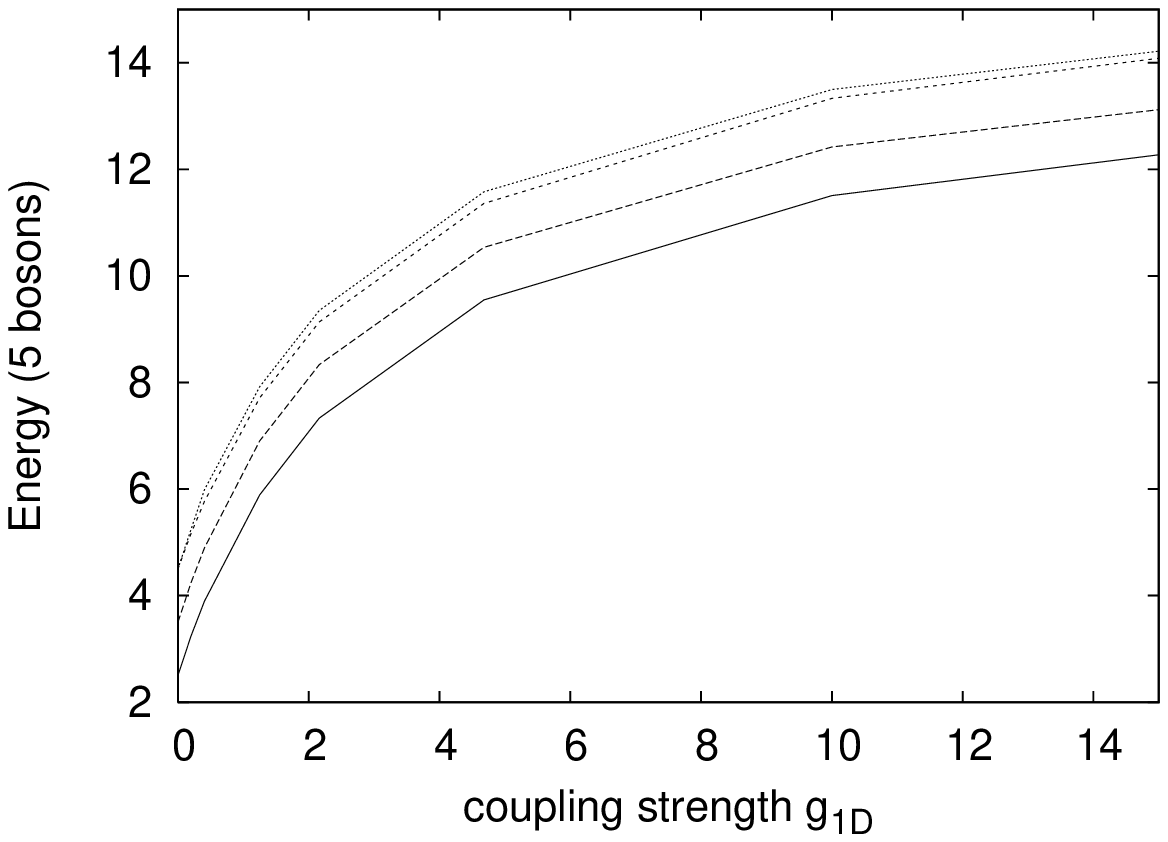}

\caption{Lowest energies $E_{m}$ in a harmonic trap ($h=0$) for $N=3,4,5$
bosons. (The lines connect the data points to guide the eye.) \label{cap:energy_h0}}
\end{figure}
For a single well, the one-particle spectrum $\left\{ \epsilon_{a}=a+{\scriptstyle \frac{1}{2}}\right\} $
is known analytically, which readily equips us with the full spectrum
for both the non-interacting and the fermionization limit. First consider
the case $g=0$. Then $E_{0}=N/2$, while all other levels follow
with an equal spacing of $\Delta_{0}=1$. Owing to that equidistance,
the degree of degeneracy $\#=E_{m}-E_{0}$ $(m\ge1)$ goes up with
each step, measured by the average occupation $N\bar{a}\equiv\sum_{a}n'_{a}a$.
Explicitly, while both $m=0,1$ are non-degenerate, the eigenspace
pertaining to $E_{2}=E_{3}=N/2+2$ is two-dimensional (see Fig.~\ref{cap:energy_h0}),
etc.

To understand this degeneracy and how it is lifted, let us recall
that, in a harmonic trap with homogeneous interactions $V(x_{i}-x_{j})$,
the center of mass (CM) $R:=\sum_{i=1}^{N}x_{i}/N$ is separable from
the relative motion. Hence one can decompose the Hilbert space $\mathbb{H=H_{\mathrm{CM}}}\otimes\mathbb{H}_{\mathrm{rel}}$
so as to write\[
\Psi=\phi_{\mathcal{N}}\otimes\psi_{\mathrm{rel}};\quad E(g)=(\mathcal{N}+{\scriptstyle \frac{1}{2}})+\epsilon_{\mathrm{rel}}(g).\]
This signifies that for every level for the \emph{relative} motion,
$\epsilon_{\mathrm{rel}}(g)$, there is a countable set of copies
shifted upward by $\mathcal{N}=0,1,\dots$ . For $g=0$, $\psi_{\mathrm{rel}}$
is a harmonic eigenstate as well, so $\epsilon_{\mathrm{rel}}^{(\nu)}(0)=\nu+{\scriptstyle \frac{N-1}{2}}$
for some $\nu$, and several different combinations of $(\mathcal{N},\nu)$
may coincide. Switching on $g>0$, however, breaks that symmetry,
leaving $\mathcal{N}$ untouched while pushing each level $\epsilon_{\mathrm{rel}}^{(\nu)}$
upward---which materializes in different slopes \[
\left.\frac{dE}{dg}\right|_{0}=\left.\frac{d}{dg}\epsilon_{\mathrm{rel}}\right|_{0}.\]
This fact is nicely illustrated on the example of $N=2$ atoms, where
\cite{Busch98} \[
\left.\frac{d}{dg}\epsilon_{\mathrm{rel}}^{(\nu)}\right|_{0}\negthickspace=\langle\psi_{\nu}|\delta(r)|\psi_{\nu}\rangle=\left|\psi_{\nu}(0)\right|^{2}\propto\binom{\nu-\frac{1}{2}}{\nu},\]
$\binom{\cdot}{\cdot}$ denoting the binomial coefficient; so higher
excited relative states {}`feel' the interaction less. This fits
in with our findings in Fig.~\ref{cap:energy_h0}: The two states
$m=2,3$ break up, the lower curve--- in light of the reasoning above---pertaining
to higher internal excitation.

Apart from that, the spectral pattern does not give an air of being
overly intricate but follows the general theme known from the two-atom
case. All levels first rise quickly in the linear perturbative regime,
but start saturating once they enter the strongly interacting domain
($g\sim10$). As insinuated, the fermionization limit is known exactly,
which endows us with a helpful calibration. Since the limits $g\to0(\infty)$
can be regarded simply as bosonic (fermionic) counterparts of the
same non-interacting system, the two should share exactly the same
energy scales (here $\Delta_{0}=1$). Indeed, building on the ground-state
energy $E_{0}=\sum_{a<N}\epsilon_{a}=N^{2}/2$, all levels again follow
in equal steps $\Delta_{0}$, with a degeneracy $\#=E_{m}-E_{0}$.
This fact, effortless as it may come out of the theory, is a strong
statement, for it implies that the very interaction that divorces
some degenerate lines at $g=0$ is also responsible for gluing them
together again if it gets sufficiently repulsive. An indication of
this effect may actually be observed in Fig.~\ref{cap:energy_h0}.

\subsubsection*{Double well ($h=5$)}

As opposed to the purely harmonic trap, the single-particle spectrum
$\{\epsilon_{a}\}$ of the double well is not that simple. In order
to get a rough idea, it is legitimate to consider a toy model of a
delta-type barrier $h\delta(x)$ (i.e., $w\to0$) \cite{Busch03}.
Then only the \emph{even} single-particle functions $\phi_{a}$ will
be affected as they have nonzero amplitudes at $x=0$. For $h>0$,
these will be notched at zero, and in the limit of large enough barriers,
$h\to\infty$, their density will approach that of the next (odd)
orbital $\phi_{a+1}$---while still remaining even---and so will their
energy, $\epsilon_{a}\to\epsilon_{a+1}$ from below. In that extreme
case, we would end up with a set of doublets separated by gaps $\Delta\epsilon=2$.
The non-interacting many-body spectrum $\{ E_{\boldsymbol{n}}=\sum_{a}n_{a}^{\prime}\epsilon_{a}\}$
would then be composed of a lowest \emph{band} of states within the
$(1+N)$-dimensional subspace $\mathrm{span}\{|n'_{0},n'_{1}=N-n'_{0}\rangle\mid n'_{0}=0,\dots N\}$,
while the next band (obtained by removing one particle from the lowest
levels $\epsilon_{0/1}$) would be shifted upward by $\Delta\epsilon$. 

\begin{figure}
\begin{center}\includegraphics[%
  width=7cm,
  keepaspectratio]{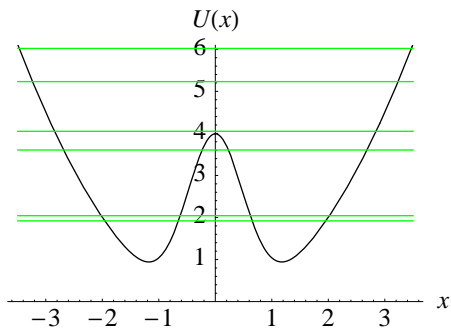}\end{center}

\caption{(color online) Single-particle spectrum $\left\{ \epsilon_{a}\right\} $
in a double well with barrier height $h=5$. \label{cap:DWlevels}}
\end{figure}
\begin{figure}
\includegraphics[%
  width=7.3cm,
  keepaspectratio]{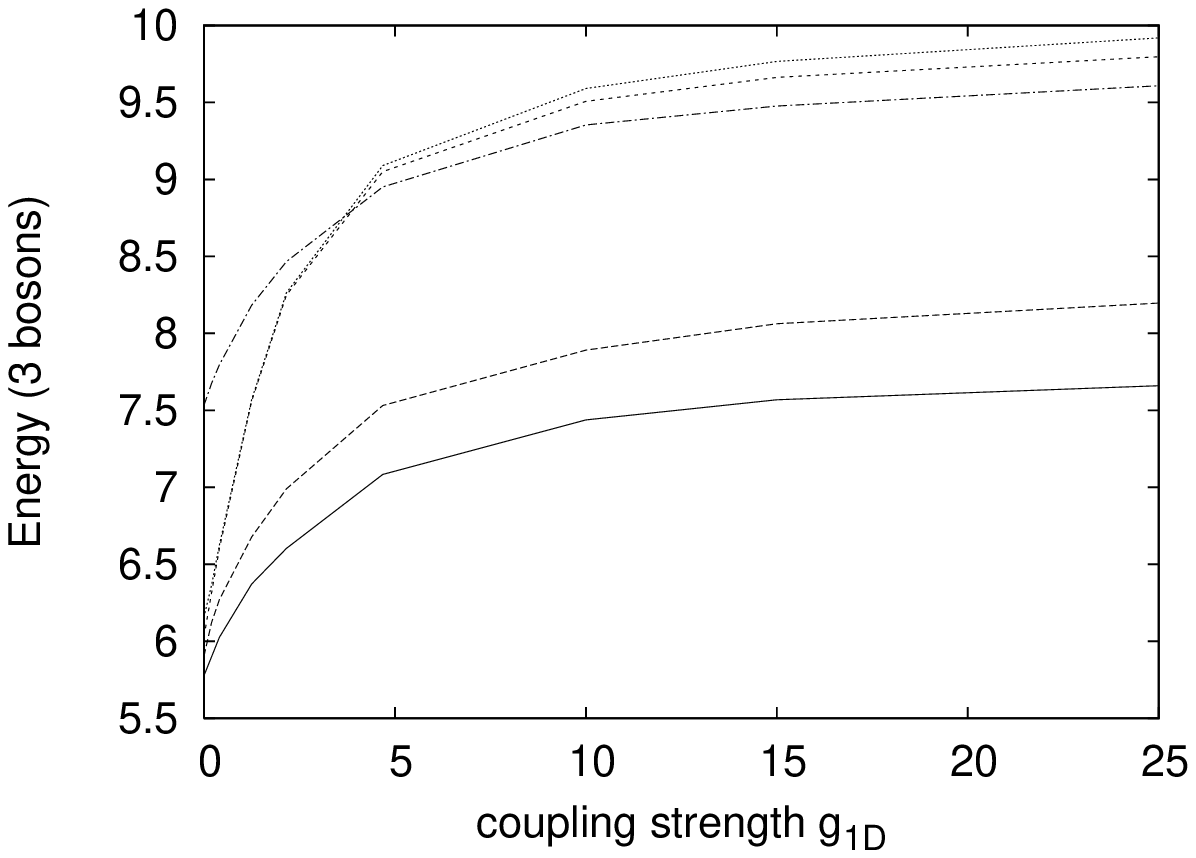}

\includegraphics[%
  width=7.3cm,
  keepaspectratio]{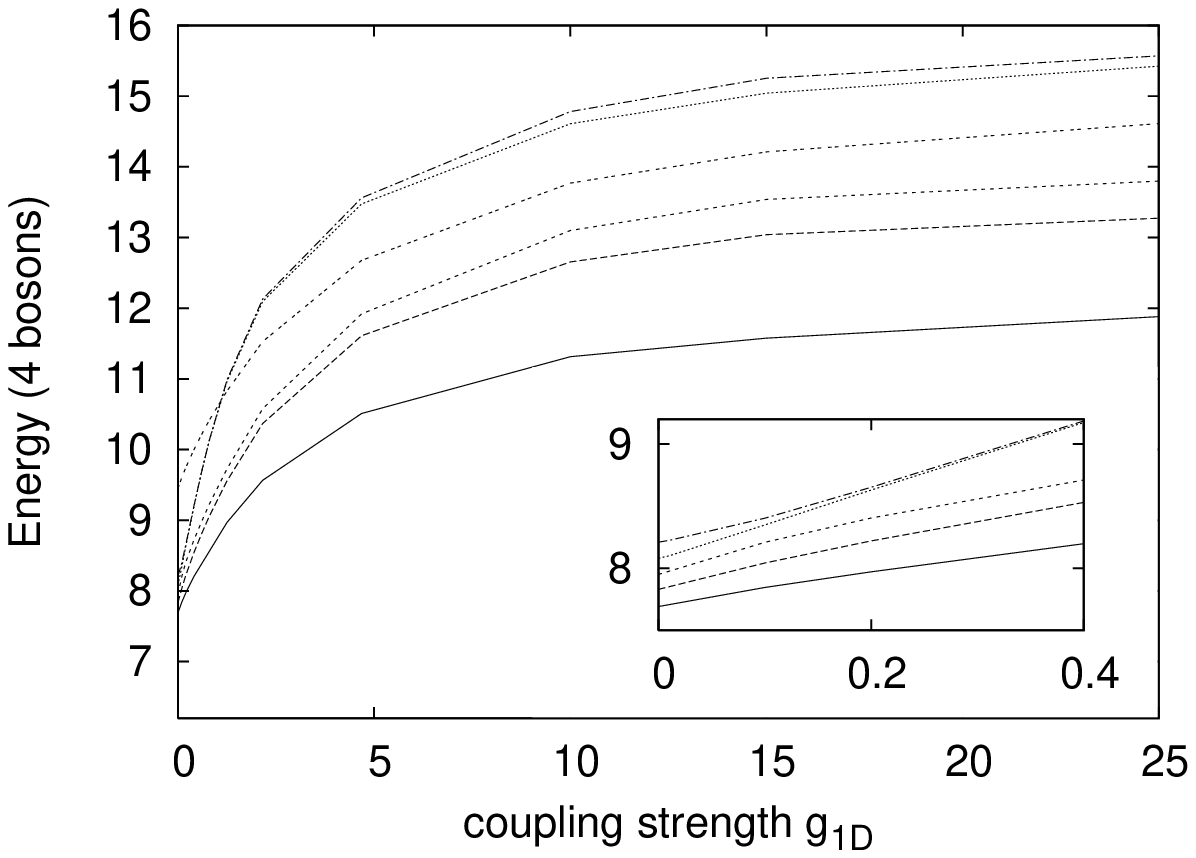}

\includegraphics[%
  width=7.3cm,
  keepaspectratio]{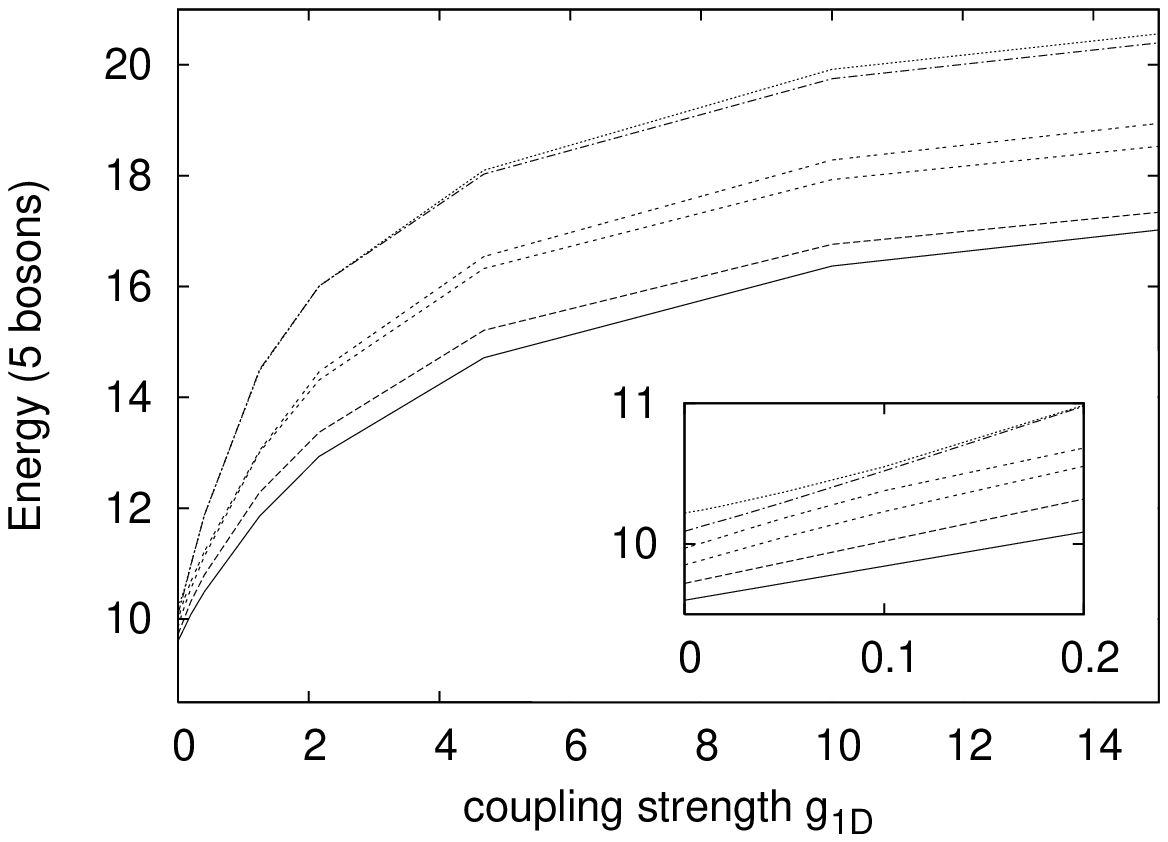}

\caption{Lowest energies $E_{m}$ in a double well ($h=5$) for $N=3,4,5$
bosons. Inset ($N=4$): level adhesion for the states $m=3,4$ (counted
from below at $g=0$). \label{cap:energy_h5}}
\end{figure}

The \emph{realistic} single-particle spectrum is sketched in Fig.~\ref{cap:DWlevels}.
Due to the nonzero width $w$, also the odd states are shifted slightly,
and of course the distance between the doublets is never exactly $\Delta\epsilon=2$
as in the crude estimate above. What is more, the finite barrier $h=5$
not only lifts the even-odd degeneracy already for the lowest states,
where the doublet character is still pronounced and the gap $\Delta_{h}\equiv\epsilon_{1}-\epsilon_{0}\ll1$,
but even more so for levels above the barrier. There the central region
is classically allowed and the spectrum becomes more and more harmonic
for higher $\epsilon_{a}$. The consequences for the \emph{full} non-interacting
spectrum $\sigma\left(H\right)$ in Fig.~\ref{cap:energy_h5} are
that the lowest band is {}`fanned up' into $\{ E_{m}=N\epsilon_{0}+m\Delta_{h}\}$.
The next cluster of levels is still well separated in energy.

The situation gets slightly more involved in the fermionization limit
$g\to\infty$. Here the spectrum is generated by (fictitious) fermionic
states $|\boldsymbol{n}\rangle$ with $n'_{a}\in\{0,1\}$, so $E_{\boldsymbol{n}}=\sum_{a}n_{a}'\epsilon_{a}$
is obtained by removing particles from below the Fermi edge $\epsilon_{N}=\mu_{N}$.
The first excited state would thus be higher in energy by $\epsilon_{N}-\epsilon_{N-1}$,
which---in our simplistic model invoked before---would be about zero
for \emph{odd} $N$ and two for \emph{even} $N$. The next band would
then be made up of four levels for any number of atoms, created by
pushing one particle out of the doublet pertaining to $\epsilon_{N-1}$
into the next upper doublet. However, since these involve higher energies,
the typical doublet structure encountered in the lowest levels is
lost for our \emph{finite} barrier $h=5$, and the spectrum will be
rearranged and mixed, as seen in Fig.~\ref{cap:energy_h5}. Here
the bands are no longer separated but smeared out considerably. Still,
the odd-vs-even distinction with respect to the lowest level may be
observed.

How do these two ends of the spectrum connect? As can be seen in Fig.~\ref{cap:energy_h5},
the reordering that is necessitated by the Bose-Fermi map starts taking
place already for small $g$, where the influence of the interaction
operator $H_{\mathrm{I}}:=\sum_{i<j}V(x_{i}-x_{j})$ can still be
treated as a perturbation,\[
\Delta E_{\boldsymbol{n}}(g)\sim\langle H_{\mathrm{I}}\rangle_{\boldsymbol{n}}+\sum_{\boldsymbol{n}'\neq\boldsymbol{n}}\frac{\left|\langle\boldsymbol{n}'|H_{\mathrm{I}}|\boldsymbol{n}\rangle\right|^{2}}{E_{\boldsymbol{n}}(0)-E_{\boldsymbol{n}'}(0)}.\]
In light of this, the different effects of $H_{\mathrm{I}}$ on different
(non-interacting) states $|\boldsymbol{n}\rangle$ will manifest themselves
in (i) their slopes and (ii) different curvatures. While the slopes
$E'(0)>0$ do not differ dramatically, the second order fosters level
repulsion within each band, since the energy gaps are small in magnitude
($\left|E_{\boldsymbol{n}}(0)-E_{\boldsymbol{n}'}(0)\right|=O(\Delta_{h=5})\ll1$)
but have different signs. (Needless to say, by conservation of parity
$\Pi\Psi(Q)\equiv\Psi(-Q)$ solely states within the same symmetry
subspace $\{\Psi\mid\Pi\Psi=\pm\Psi\}$ are coupled.) Even though
limited to perturbative values $g<1$, this may help us understand
why the upper lines of the lowest band tend to rise so steeply, whereas
the lowest ones in each band are {}`pushed' downward a little. For
example, see Fig.~\ref{cap:energy_h5}, $N=4$: the two levels $m=3,4$
(counted at $g=0$) intersect the lowest upper-band state $|3,0,1\rangle$
at $g\simeq1$.

Most stunning is the observation that, apparently, some lines are
virtually {}`glued together' once the interaction is switched on
(see Fig.~\ref{cap:energy_h5}; insets $N=4,5$). In particular,
this applies to the upper pair $m=N-1,N$ (counted at $g=0$, as always).
To be sure, the two levels are close from the start ($\Delta E=\Delta_{5}\simeq0.1$);
but for $g\sim1$ they get as close as $\sim0.01$. This quasi-degeneracy
arises as interactions are turned on; but it is destroyed again once
these get very strong ($g\sim5$). However, there is no indication
as to what exact mechanism brings the two lines together. Not only
do the corresponding states remain orthogonal at all couplings, but
they stem from opposite-parity subspaces and can thus only mix with
other states of a kind. Still, their reduced densities will turn out
very much alike (as will be laid out in Sec.~\ref{sub:states}).

\subsection{Excited states \label{sub:states}}

As yet, we have looked into the spectrum and its evolution from the
weakly to the strongly interacting regime. In Sec.~\ref{sub:spectrum},
we have found characteristic spectral patterns for the cases of a
single and a double well, respectively. We now aspire to get a deeper
insight into the underlying states $\Psi_{m\ge1}$, which may be also
beneficial for studying the dynamics in future applications.

Generally speaking, the non-interacting limit is described in terms
of number states $|\boldsymbol{n}\rangle$ in the respective one-particle
basis. Owing to the asymptotically harmonic confinement, we thus have
an overall Gaussian profile $\rho(x)\propto\exp\left(-x^{2}\right)$,
which is modulated by the central barrier as well as the degree of
excitation. At least for the low-lying states, the length scale is
therefore about that of the harmonic confinement, $a_{\parallel}=1$.
Being single-particle states, they are essentially devoid of two-body
correlations, which  amount to $\rho_{2}=\frac{1}{2}(1+P_{12})\rho_{1}\otimes\rho_{1}$
(with the permutation operator $P_{12}$).

When interactions are added, some extra interaction energy $\frac{N(N-1)}{2}\mathrm{tr}\left(V\rho_{2}\right)$
must be paid. Hence, the system will respond by depleting the correlation
diagonal $\rho_{2}(x_{1},x_{2}=x_{1}$), roughly speaking. In our
single-particle expansion, this will be done by both adapting the
single-particle functions $\varphi_{j}$ as well as by mixing different
configurations $\Phi_{J}$ or---in the symmetrized version---$|\boldsymbol{n}\rangle$,
depending on how close in energy they are. As $g\to\infty$, this
culminates in the system's fermionization, where there is a one-to-one
correspondence to fermionic states, $n'_{a}=0,1$. In particular,
the density profile $\rho=\sum_{a}\left|\phi_{a}\right|^{2}$ becomes
broader, with a length scale of order $\sqrt{2N}$ \cite{kolomeisky00},
while the strongly correlated nature is captured in the fermionic
two-body density $\rho_{2}(x_{1},x_{2})=\left(\rho(x_{1})\rho(x_{2})-\left|\rho_{1}(x_{1},x_{2})\right|^{2}\right)/2$,
which naturally vanishes at points of collision.

\subsubsection*{Harmonic trap ($h=0$)}

\begin{figure}
\noindent \begin{center}\includegraphics[%
  width=6.1cm,
  keepaspectratio]{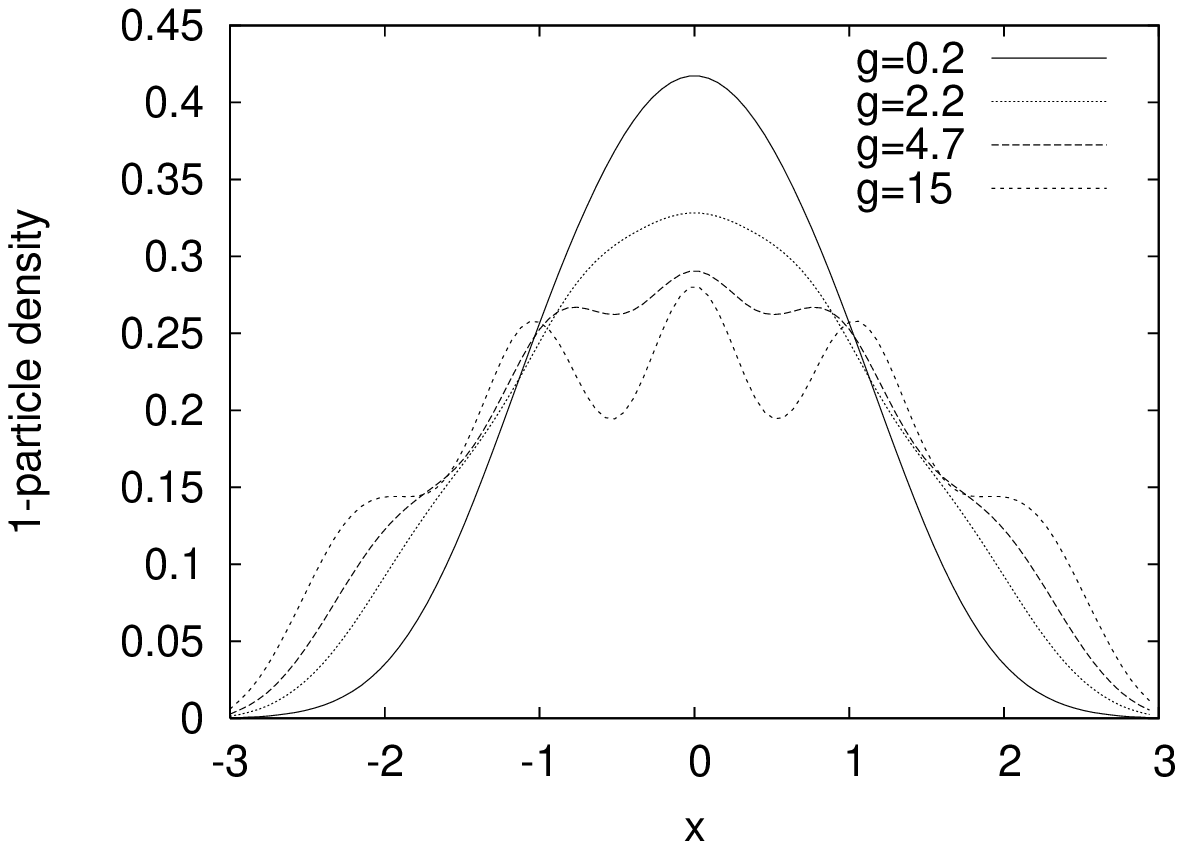}\end{center}

\noindent \begin{center}\includegraphics[%
  width=6.1cm,
  keepaspectratio]{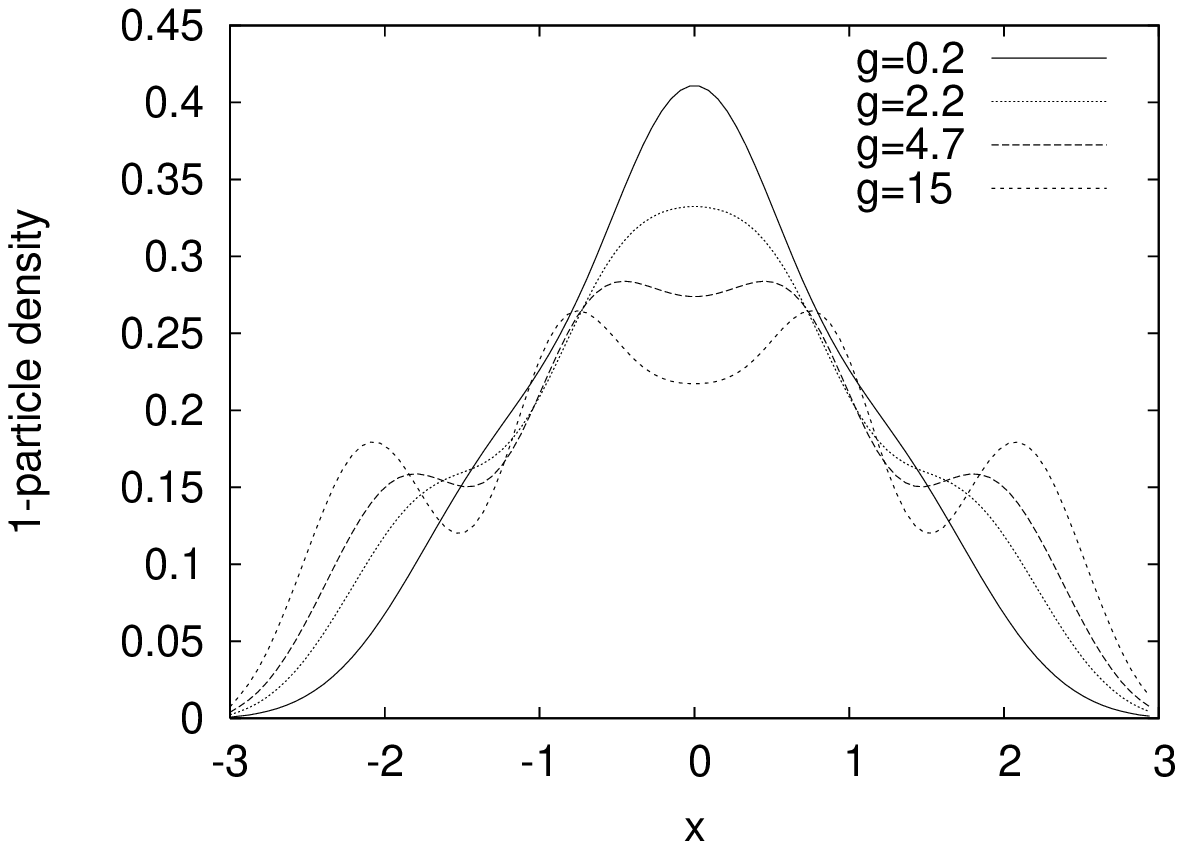}\end{center}

\noindent \begin{center}\includegraphics[%
  width=6.1cm,
  keepaspectratio]{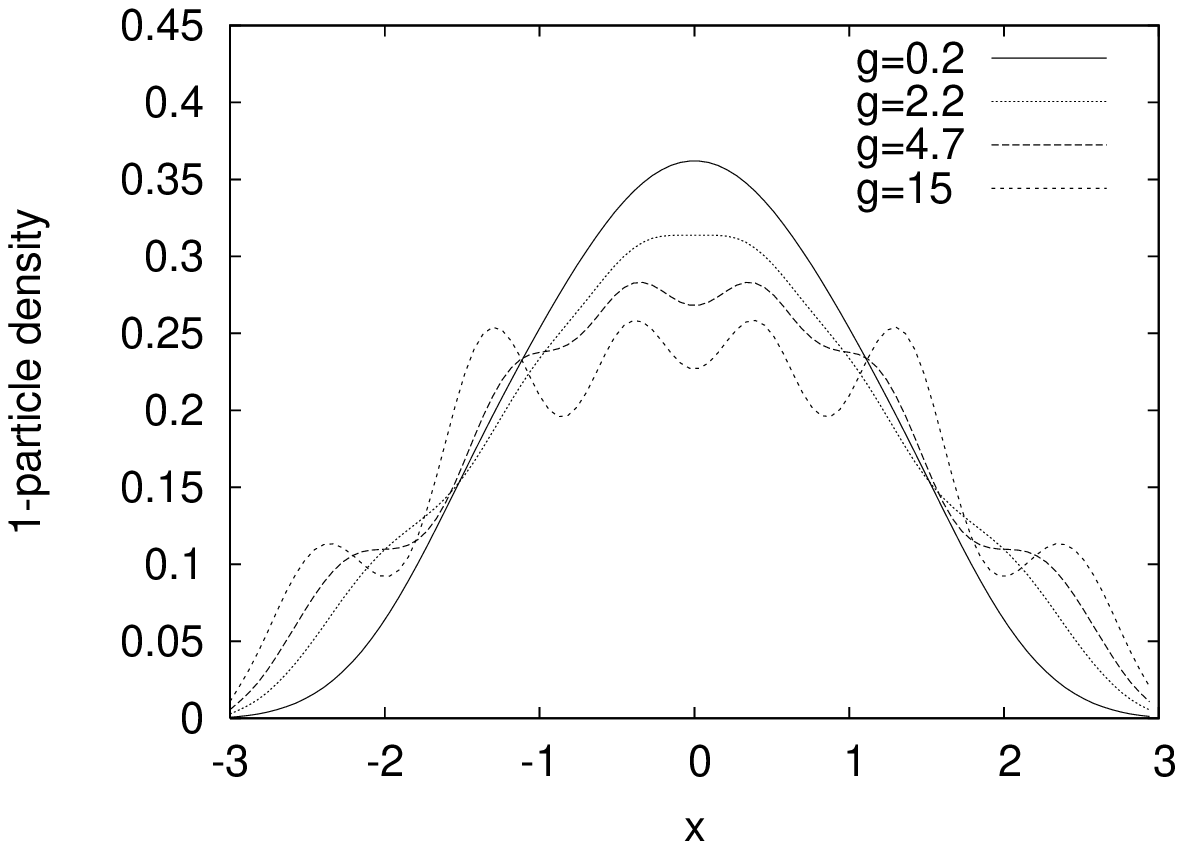}\end{center}

\caption{Density profiles of $N=4$ bosons in a harmonic trap ($h=0$) for
the excited states $m=1,2,3$ (from top to bottom). \label{cap:densities-h0} }
\end{figure}
Here the nature of the fermionization transition has been explored
extensively for the ground state \cite{deuretzbacher06,zoellner06a,zoellner06b}.
As for the excited states, a look at Fig.~\ref{cap:densities-h0}
unveils that essentially the same mechanisms are at work, as exemplified
on the one-body density $\rho(x)$ for different states $m$. The
non-interacting density profiles have a Gaussian envelope. This may
be seen in the plot for $g=0.2$, the characteristic shape for the
degenerate states $m=2,3$ stemming from the fact that the interaction
$H_{\mathrm{I}}$ selects a linear combination so as to be diagonal
within that subspace. Upon increasing $g$, the density is being flattened,
reflecting the atoms' repelling one another. Eventually, a fermionized
state is reached, featuring characteristic humps in the density. As
in the ground-state case, these signify \emph{localization}; in other
words, it is more likely to find one atom at discrete spots $x_{i}$.
However, here the fermionization pattern eludes an obvious interpretation,
since these are excited states. In particular, now the number of humps
need not equal $N$, as can be seen for $m=1$. 

\begin{figure}
\includegraphics[%
  width=3.1cm,
  keepaspectratio]{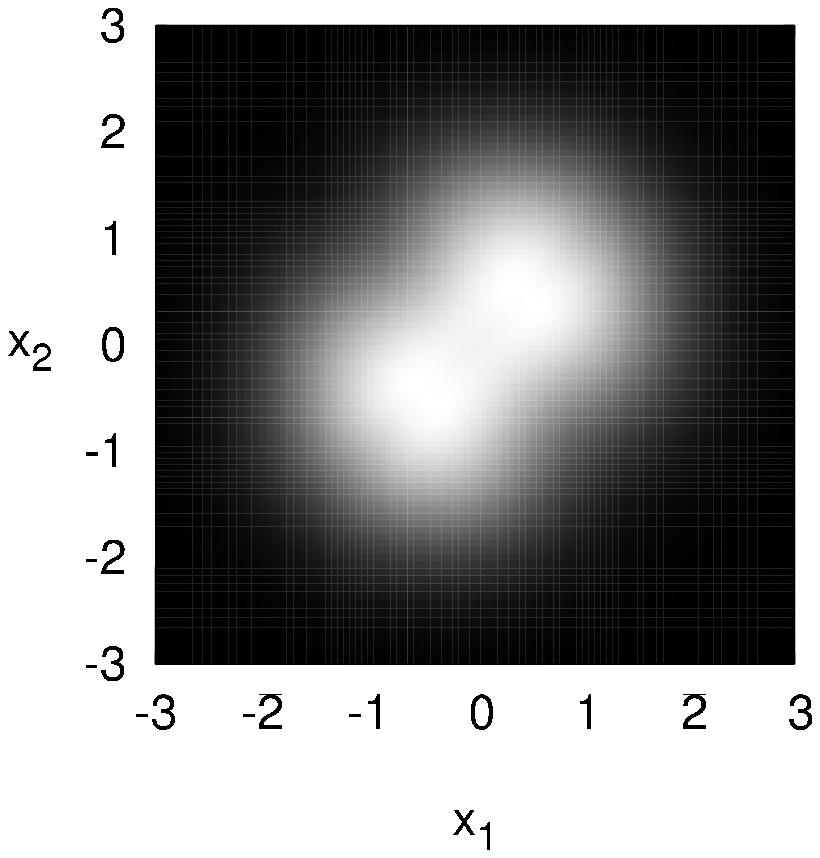}\includegraphics[%
  width=3.1cm,
  keepaspectratio]{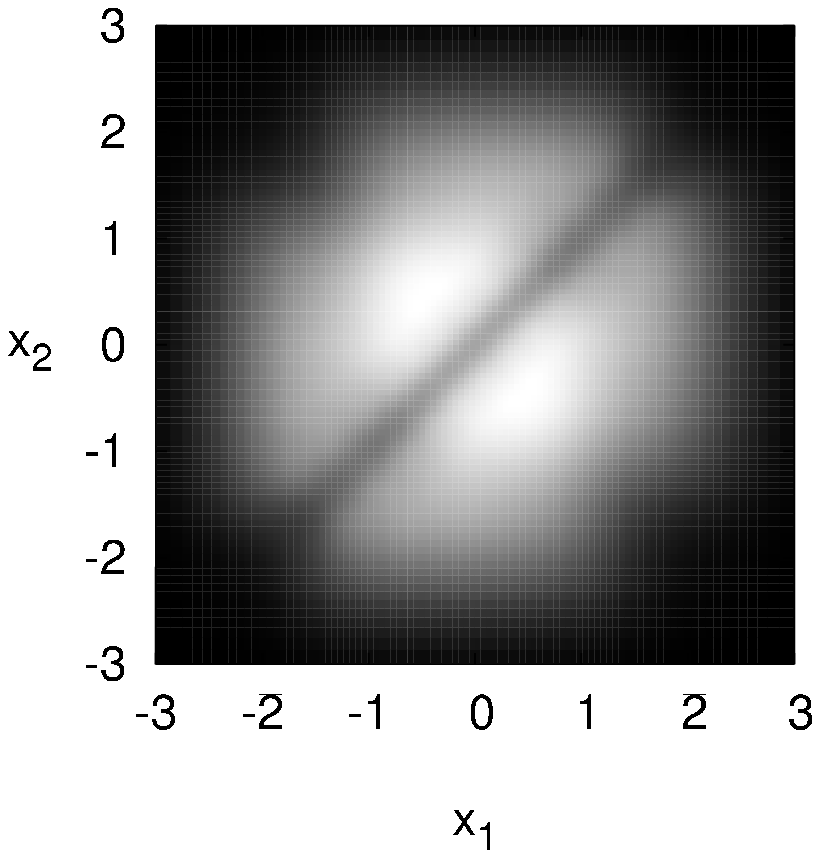}\includegraphics[%
  width=3.1cm,
  keepaspectratio]{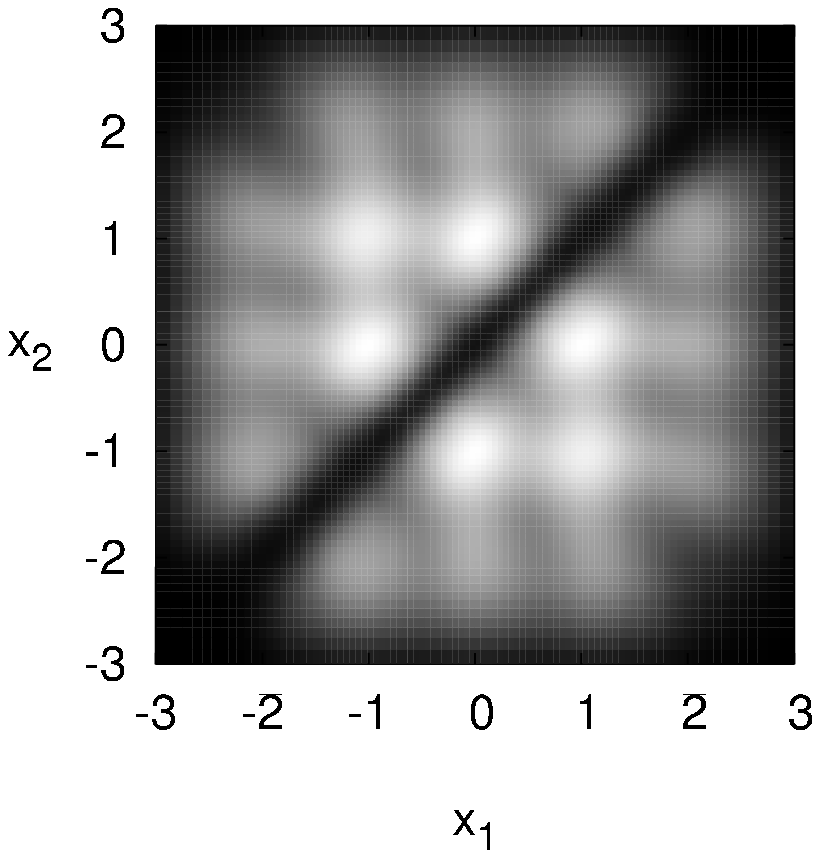}

\includegraphics[%
  width=3.1cm,
  keepaspectratio]{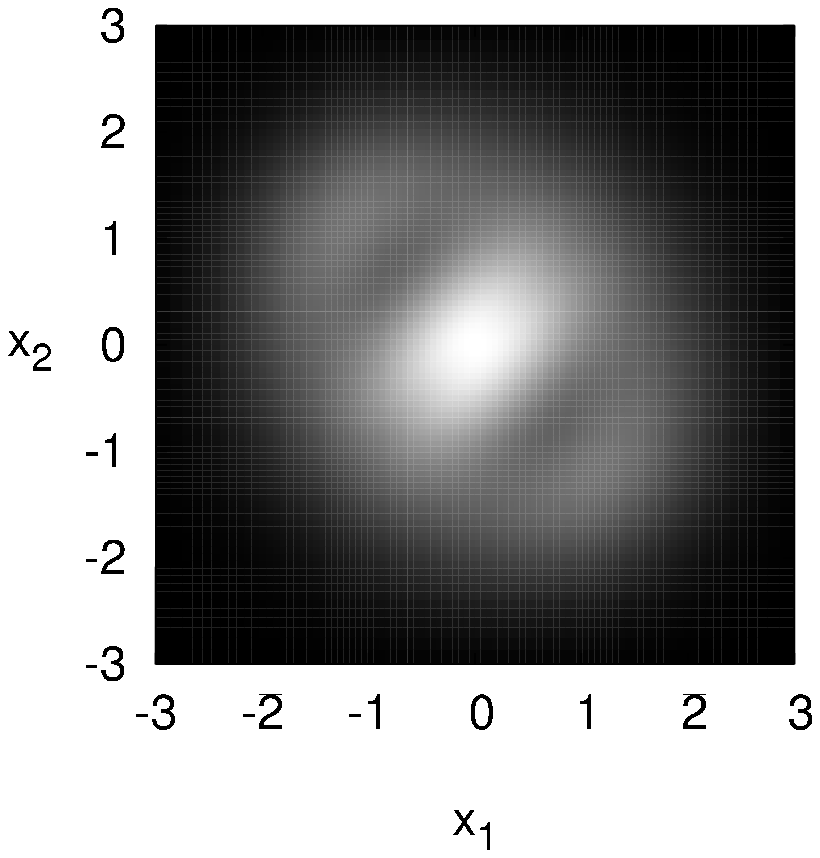}\includegraphics[%
  width=3.1cm,
  keepaspectratio]{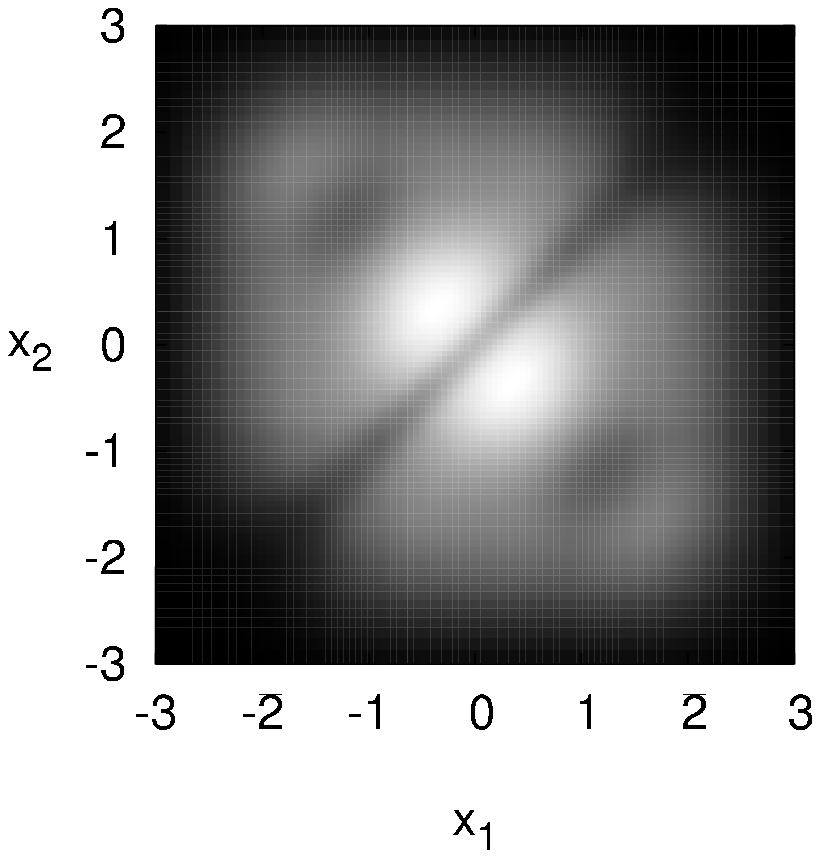}\includegraphics[%
  width=3.1cm,
  keepaspectratio]{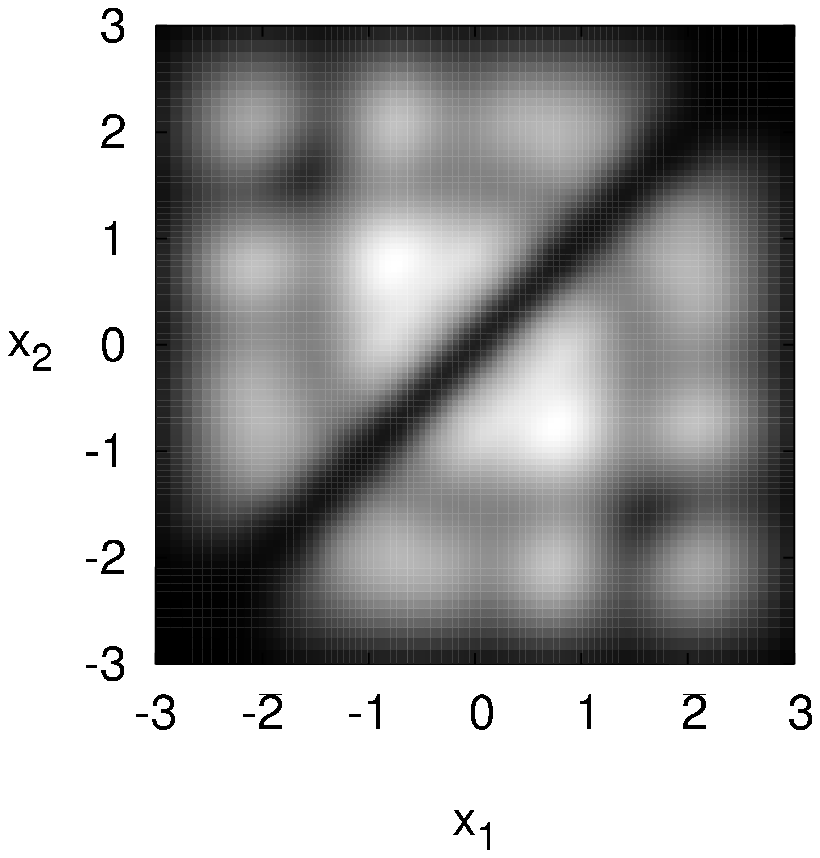}

\includegraphics[%
  width=3.1cm,
  keepaspectratio]{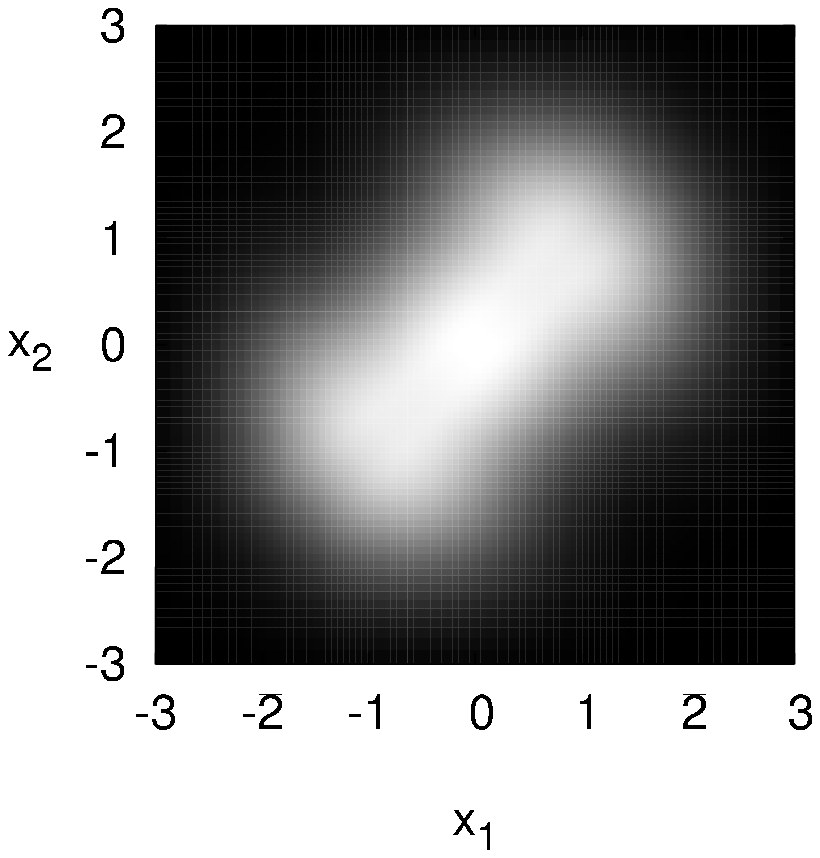}\includegraphics[%
  width=3.1cm,
  keepaspectratio]{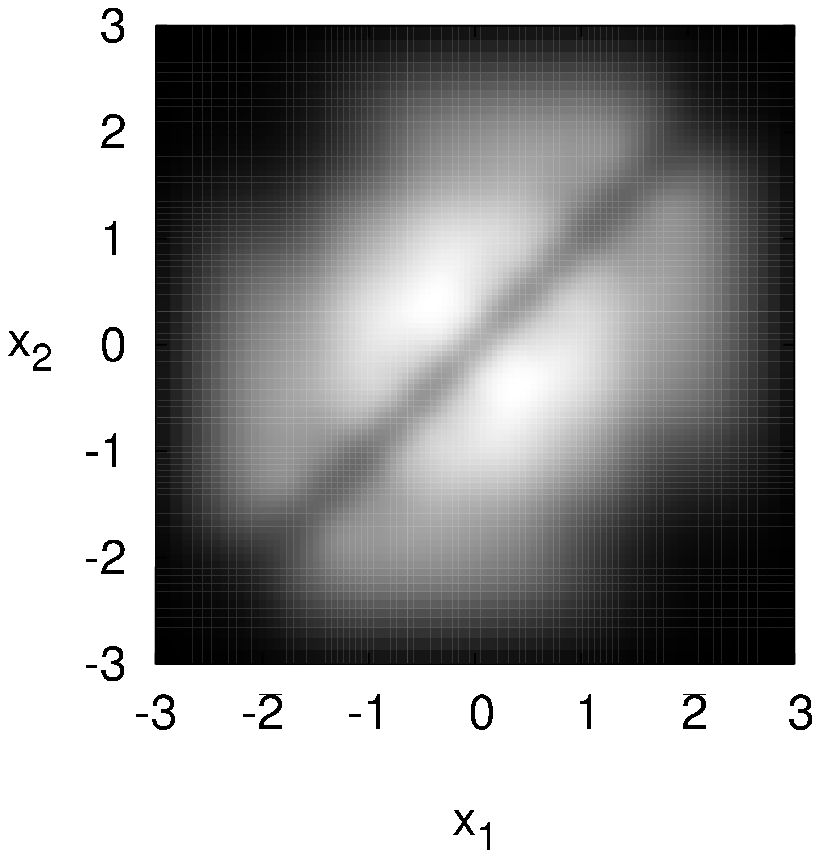}\includegraphics[%
  width=3.1cm,
  keepaspectratio]{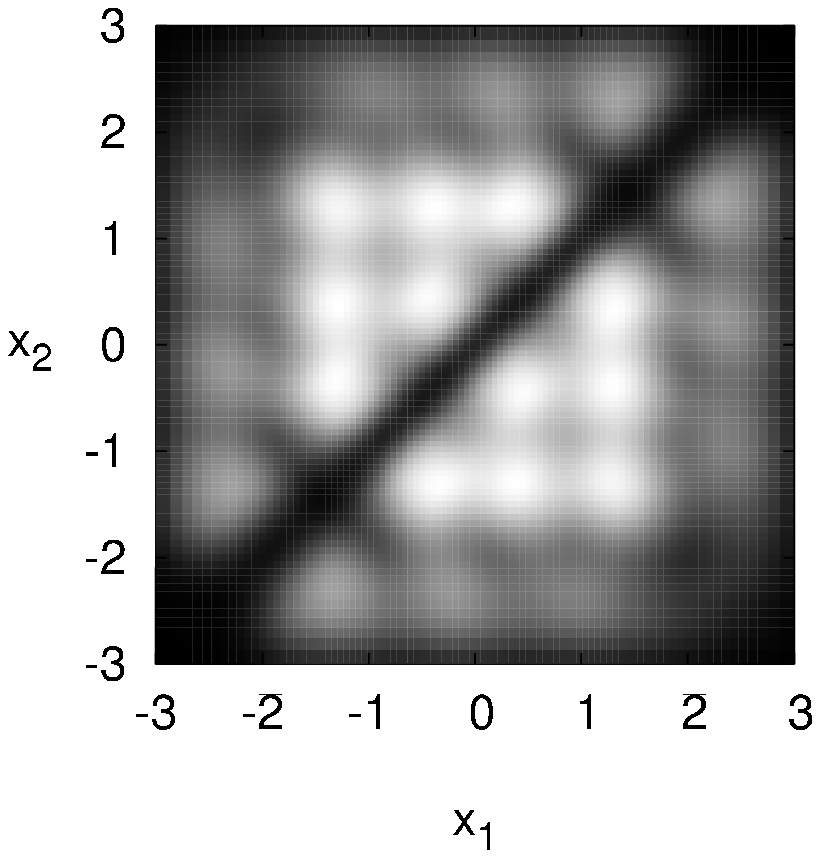}

\caption{Two-particle density $\rho_{2}(x_{1},x_{2})$ for $N=4$ bosons in
a harmonic trap. Top row: excited state $m=1,\dots$, bottom row:
$m=3$; shown are the interaction strengths $g=0.2,\,2.2,\,15$ from
left to right. \label{cap:densities2_h0}}
\end{figure}
A look behind the scene is offered by the two-body correlation function
$\rho_{2}$ (Fig.~\ref{cap:densities2_h0}), which includes $\rho=\int dx_{2}\rho_{2}(\cdot,x_{2})$
by averaging over the second atom. It illustrates nicely how the interaction
imprints a correlation hole at $\left\{ x_{1}=x_{2}\right\} $, which
relates to the washed-out profile in Fig.~\ref{cap:densities-h0}.
A complex fragmentation of the $(x_{1},x_{2})$ plane can be witnessed
as we go to larger $g$, which is different from the very obvious
checkerboard pattern of the ground-state case \cite{zoellner06a}.
The latter one provided a simple interpretation, namely that the atoms
are evenly distributed at discrete positions over the trap (up to
a Gaussian density modulation), but with zero probability of finding
two atoms at the same spot. Here the atoms are apparently more localized
in the center. On top of that, if one atoms is fixed at some $x_{1}$,
one cannot unconditionally ascribe definite positions for the $N-1$
remaining particles as before.

\subsubsection*{Double well ($h=5$)}

\begin{figure}
\includegraphics[%
  width=6.5cm,
  keepaspectratio]{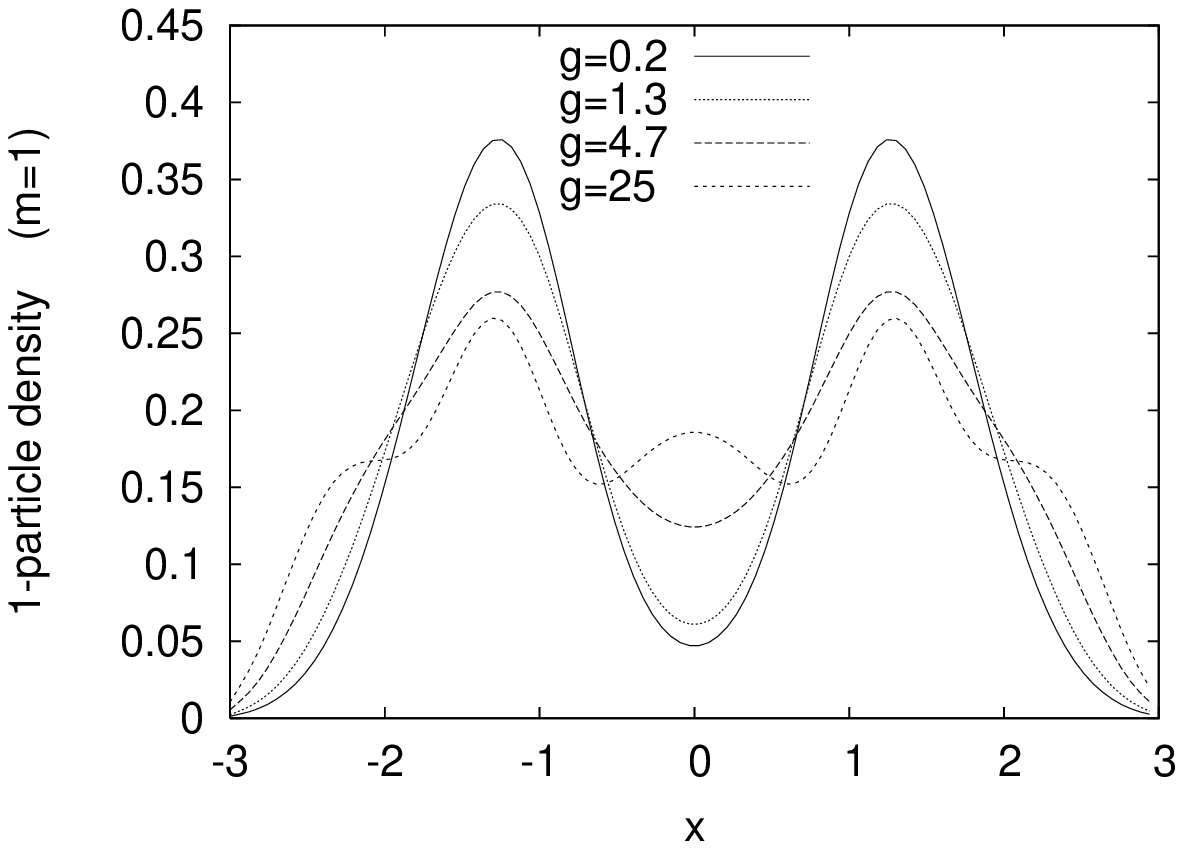}

\includegraphics[%
  width=6.5cm,
  keepaspectratio]{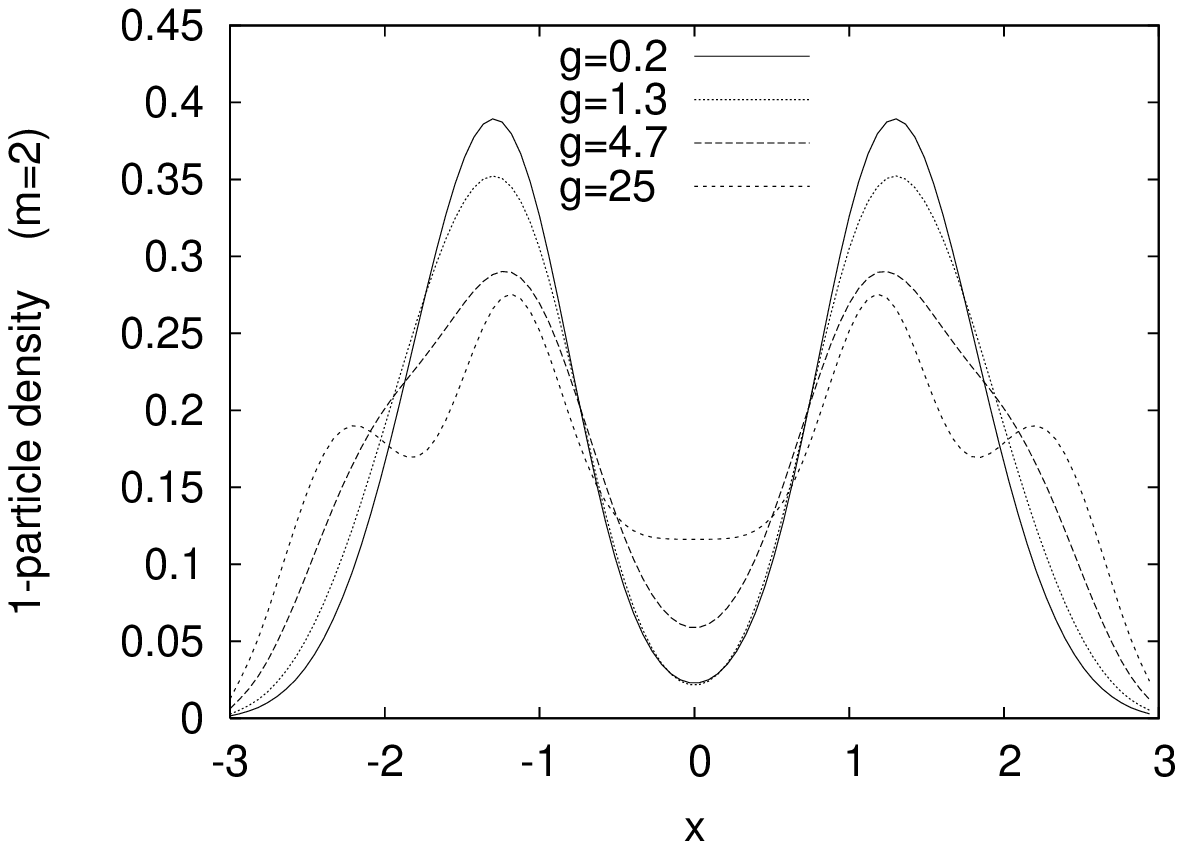}

\includegraphics[%
  width=6.5cm,
  keepaspectratio]{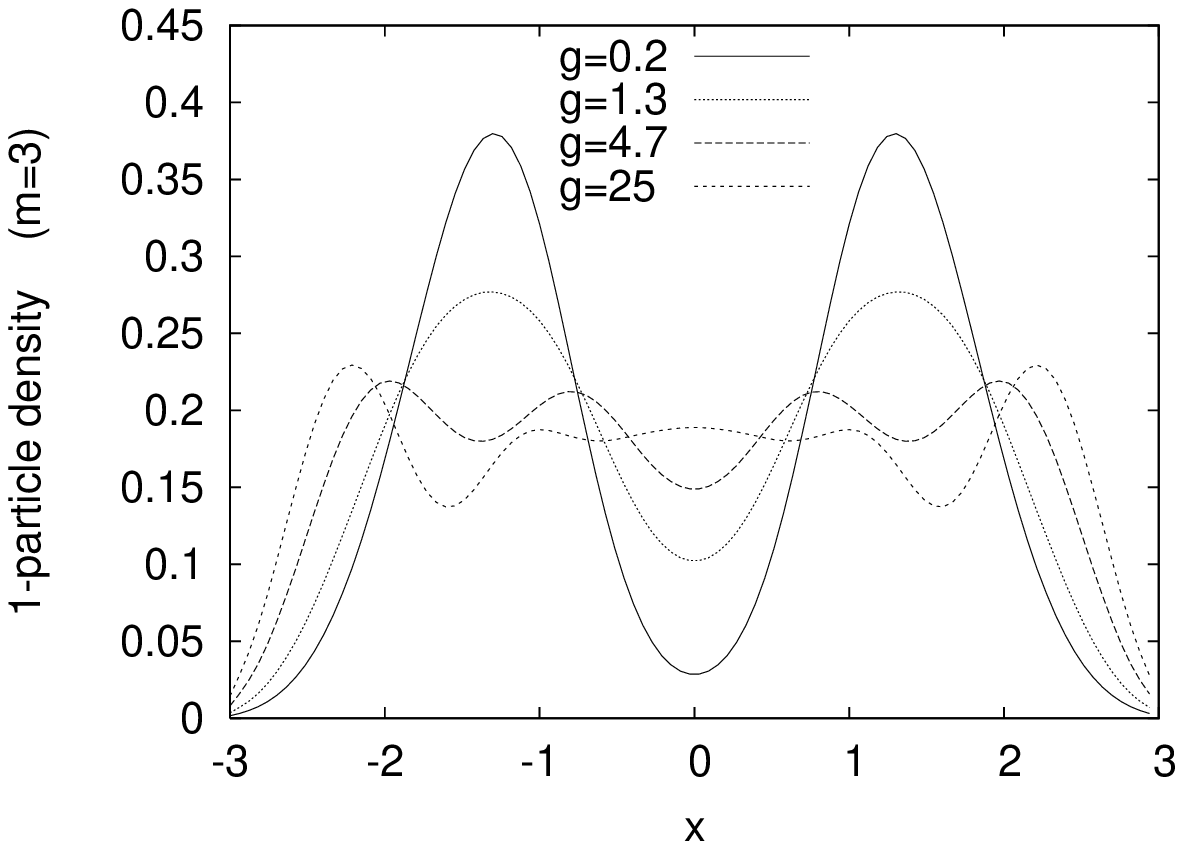}

\includegraphics[%
  width=6.5cm,
  keepaspectratio]{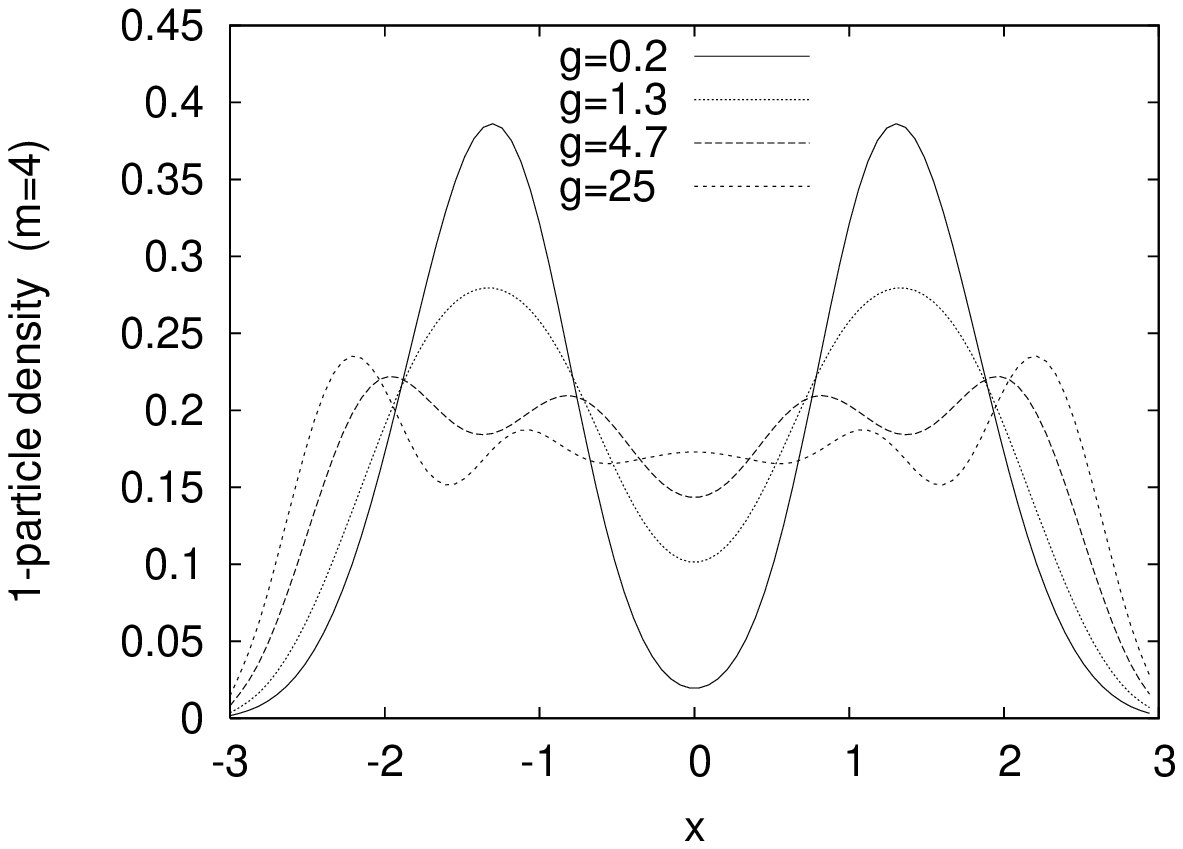}

\caption{Density profiles of $N=4$ bosons in a double well ($h=5$) for the
lowest excited states $m=1,\dots,4$ (from top to bottom). \label{cap:densities-h5} }
\end{figure}
For large but finite barrier heights, the lowest excitations at $g=0$
will be formed by the two-mode vectors $|n'_{0},N-n'_{0}\rangle$.
All of these will exhibit similar density profiles since $\rho(x)$
only differs significantly near the trap's center; specifically $\rho(0)=n_{0}\left|\phi_{0}(0)\right|^{2}$.
This can be seen in Fig.~\ref{cap:densities-h5}, which summarizes
the evolution of the lowest excited states' densities for $N=4$.
As the perturbation $H_{\mathrm{I}}$ is turned on, different neighboring
states $|\boldsymbol{n}\rangle$ (of equal symmetry) will be admixed,
and the profiles are adjusted accordingly. Eventually, they saturate
in the fermionic limit, featuring a typically broadened shape. Note
that, for intermediate interactions (e.g., $g=1.3$), the aforementioned
quasi-degenerate states $m=3,4$ (cf. Fig.~\ref{cap:energy_h5})
indeed reveal an almost identical profile. 

\begin{figure}
\includegraphics[%
  width=4.2cm,
  keepaspectratio]{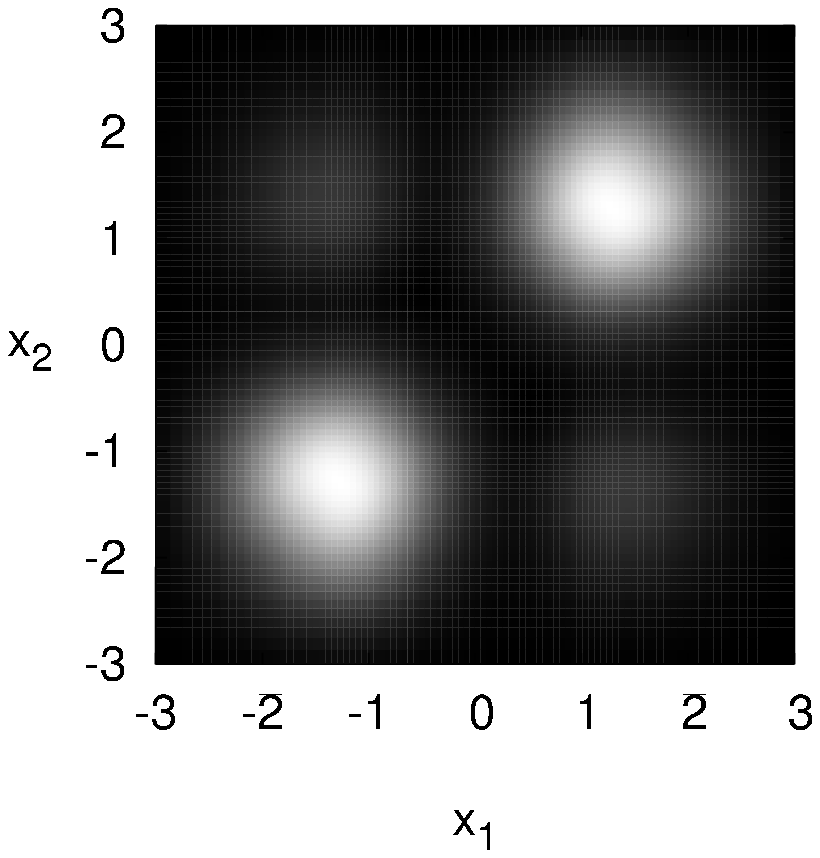}\includegraphics[%
  width=4.2cm,
  keepaspectratio]{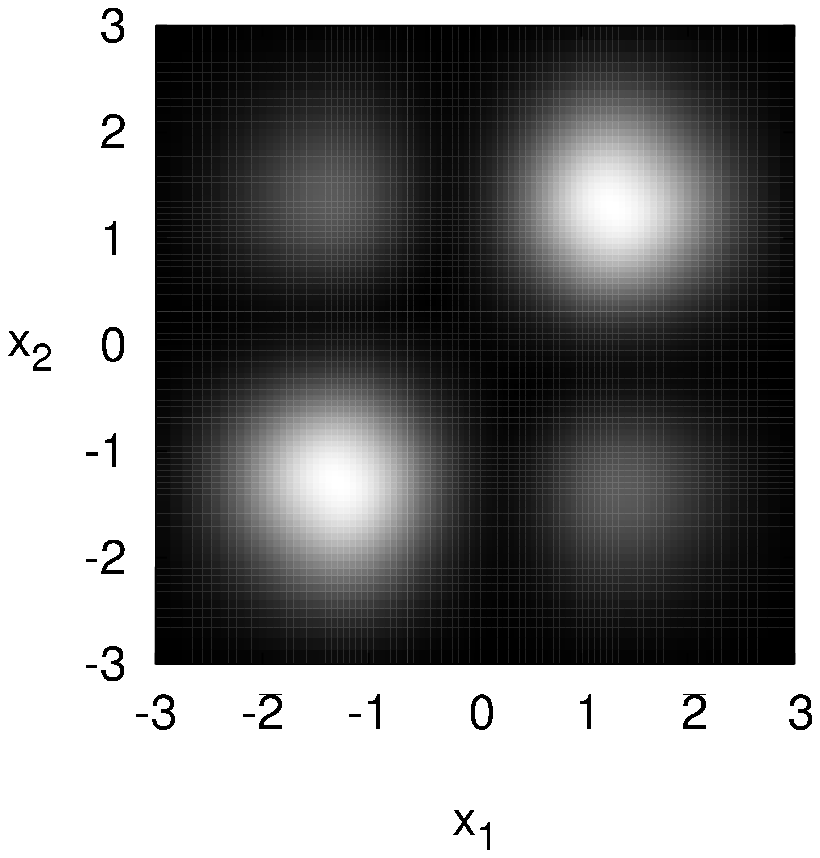}

\includegraphics[%
  width=4.2cm,
  keepaspectratio]{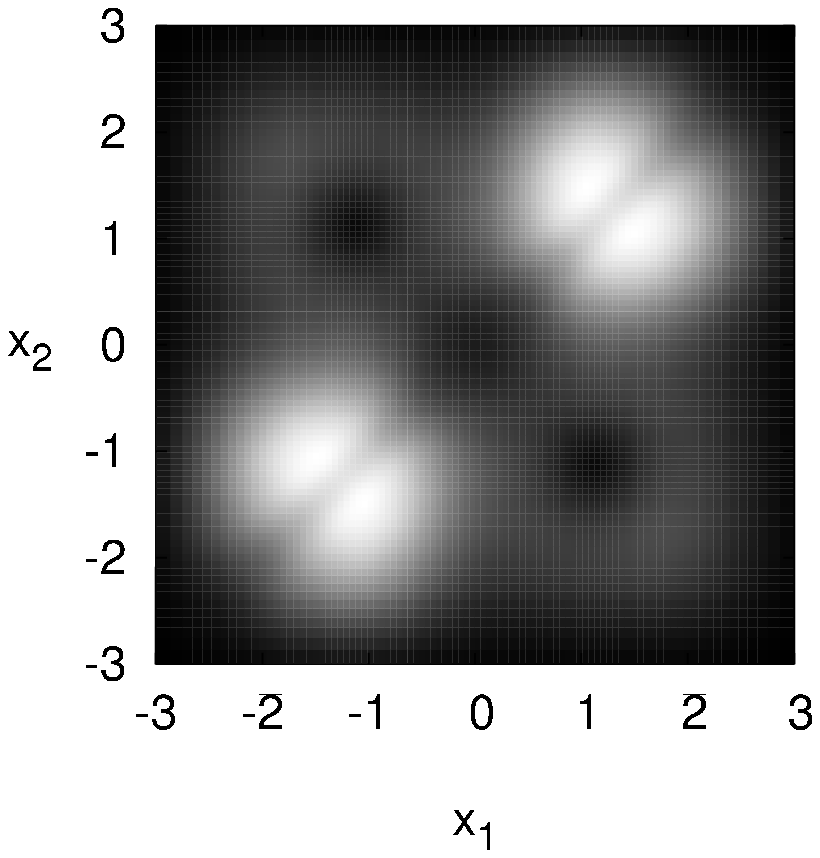}\includegraphics[%
  width=4.2cm,
  keepaspectratio]{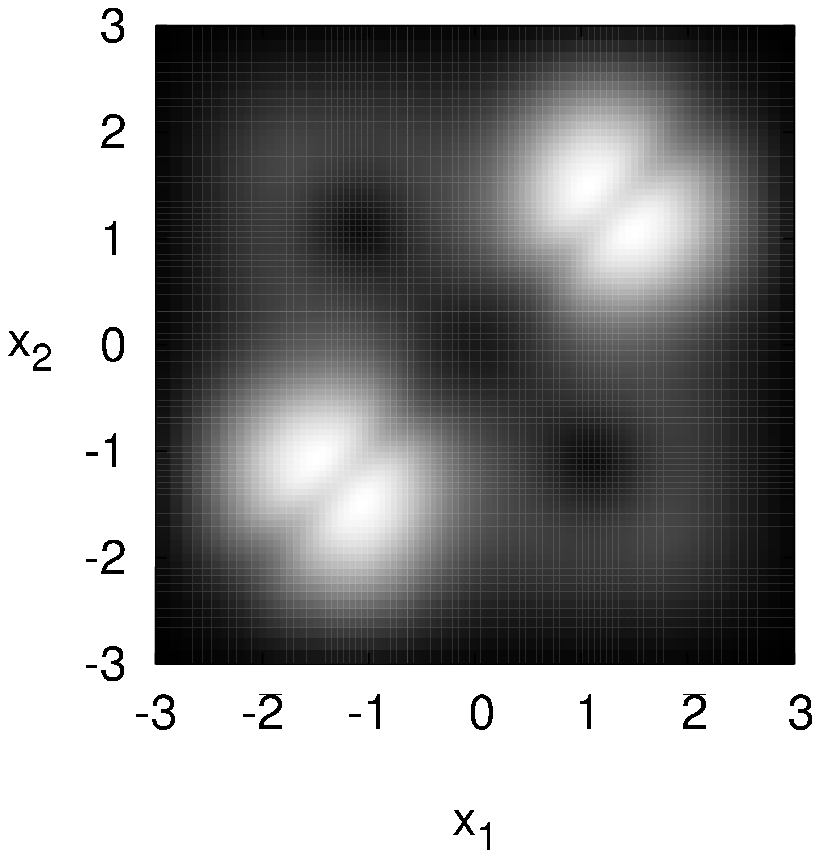}

\includegraphics[%
  width=4.2cm,
  keepaspectratio]{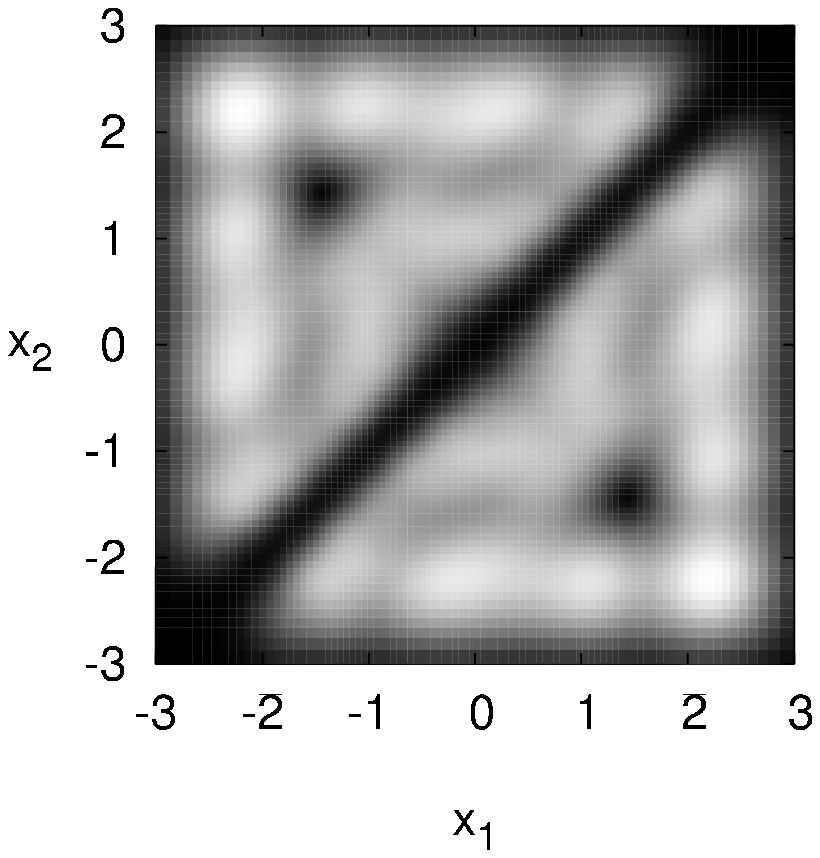}\includegraphics[%
  width=4.2cm,
  keepaspectratio]{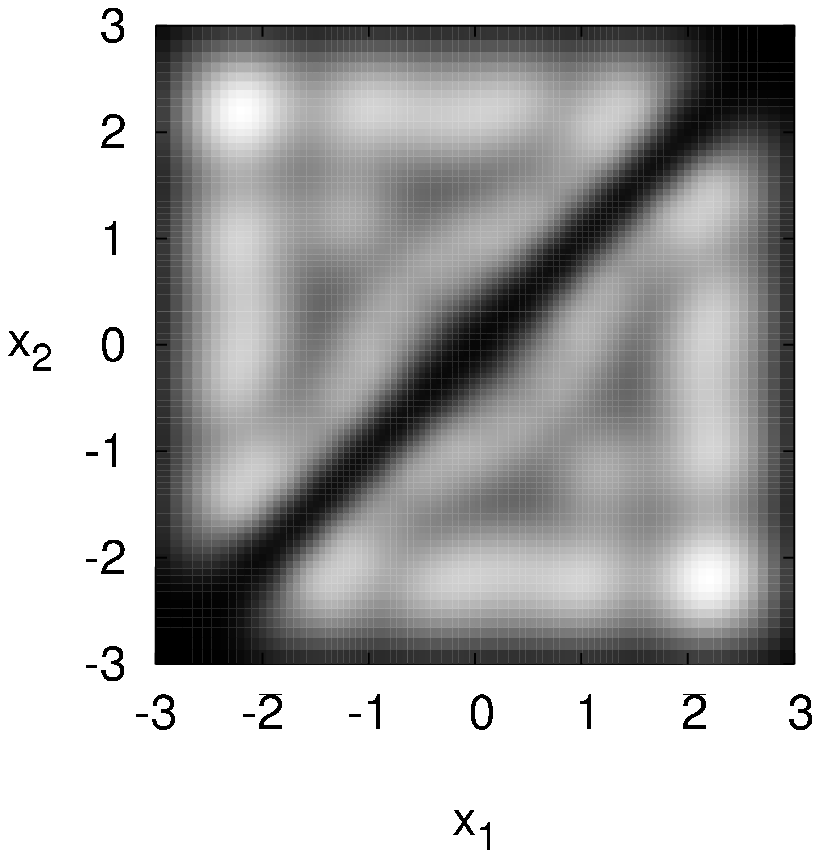}

\caption{Two-particle density $\rho_{2}(x_{1},x_{2})$ for $N=4$ bosons in
a double-well trap $(h=5)$. Left: excited state $m=3$, right: $m=4$;
shown are the interaction strengths $g=0.2,\,1.3,\,25$ from top to
bottom. The densities for both states are practically indistinguishable
for mediate $g\sim1$. \label{cap:densities2_h5}}
\end{figure}
\begin{figure}
\begin{center}\includegraphics[%
  width=7cm]{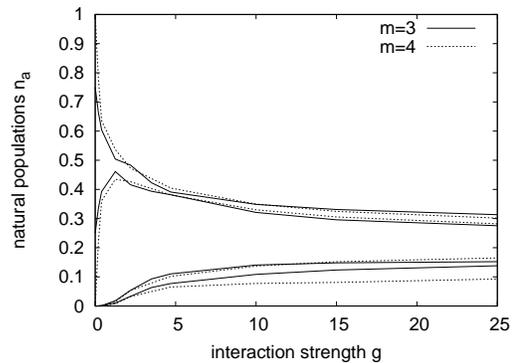}\end{center}

\caption{Natural populations $n_{a}(g)$ ($a=0,\dots,3$ from top) for the
excited states $m=3$ (---) and $m=4$ ($\cdots\cdots$) of $N=4$
bosons in a double well with barrier height $h=5$. Both are practically
equal in energy and densities for mediate $g$. \label{cap:DW.h5_natpop}}
\end{figure}

The same goes for the two-body density displayed in Fig.~\ref{cap:densities2_h5}
for these two states ($m=3,4$). Absent any interactions, $\Psi_{4}$
stands out as it has pronounced off-diagonal peaks $\rho_{2}(x,-x)=\rho_{2}(x,x)=\left|\phi_{1}(x)\right|^{4}$,
in contrast to $\Psi_{3}$. Already for small $g$, these are washed
out due to admixing of neighboring states. For intermediate couplings,
$g=1.3$, both densities become virtually indistinguishable. Eventually,
also here the fermionization limit is reached, and the densities can
be discriminated again. That said, it should once more be stressed
that the respective states are by no means similar. Rather, they are
described roughly through the orthogonal subspaces spanned by $\{|3,1\rangle,|1,3\rangle\}$
($m=3$) and $\{|2,2\rangle,|0,4\rangle\}$ ($m=4$). Figure~\ref{cap:DW.h5_natpop}
sheds light on this aspect by laying out the evolution of the natural
populations $n_{a}(g)$. It is similar for $m=3,4$ but not identical.
The residual weights $n_{a\ge2}$ are slightly separated from the
dominant ones ($a=0,1$). However, as evidenced for the ground state
\cite{zoellner06b}, they cannot be neglected because they accumulate
densely on a logarithmic scale, mirroring the extreme one-particle
correlations imprinted in the course of fermionization.%

\subsection{Crossover from single to double well \label{sub:barrier-crossover}}

We have come a long way studying in depth the spectral properties
of a single and a double well. As opposed to the ground-state case
\cite{zoellner06a}, the link between the two is far from obvious.
In the harmonic trap, the fermionization transition was fairly tame,
while in the presence of a fixed barrier $h=5$, there not only seemed
to be a strikingly different level structure to begin with, but also
the onset of a zoo of crossings and quasi-degeneracies. On that score,
it would be desirable to get an understanding of the crossover from
a single to a double well. To this end, we will again borrow some
inspiration from the simple model of a point-split trap $h\delta(x)$
\cite{Busch03}.

\begin{figure}
\begin{center}\includegraphics[%
  width=6.5cm,
  keepaspectratio]{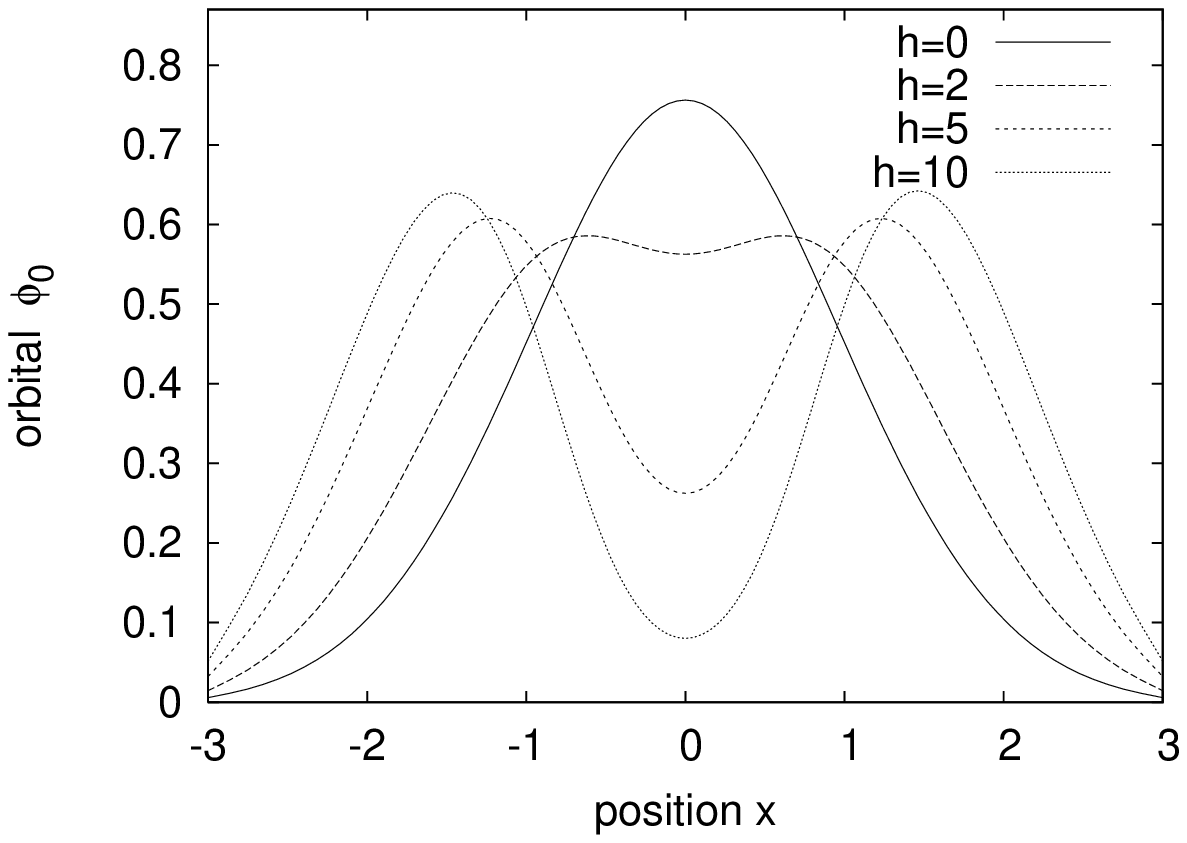}\end{center}

\begin{center}\includegraphics[%
  width=6.5cm,
  keepaspectratio]{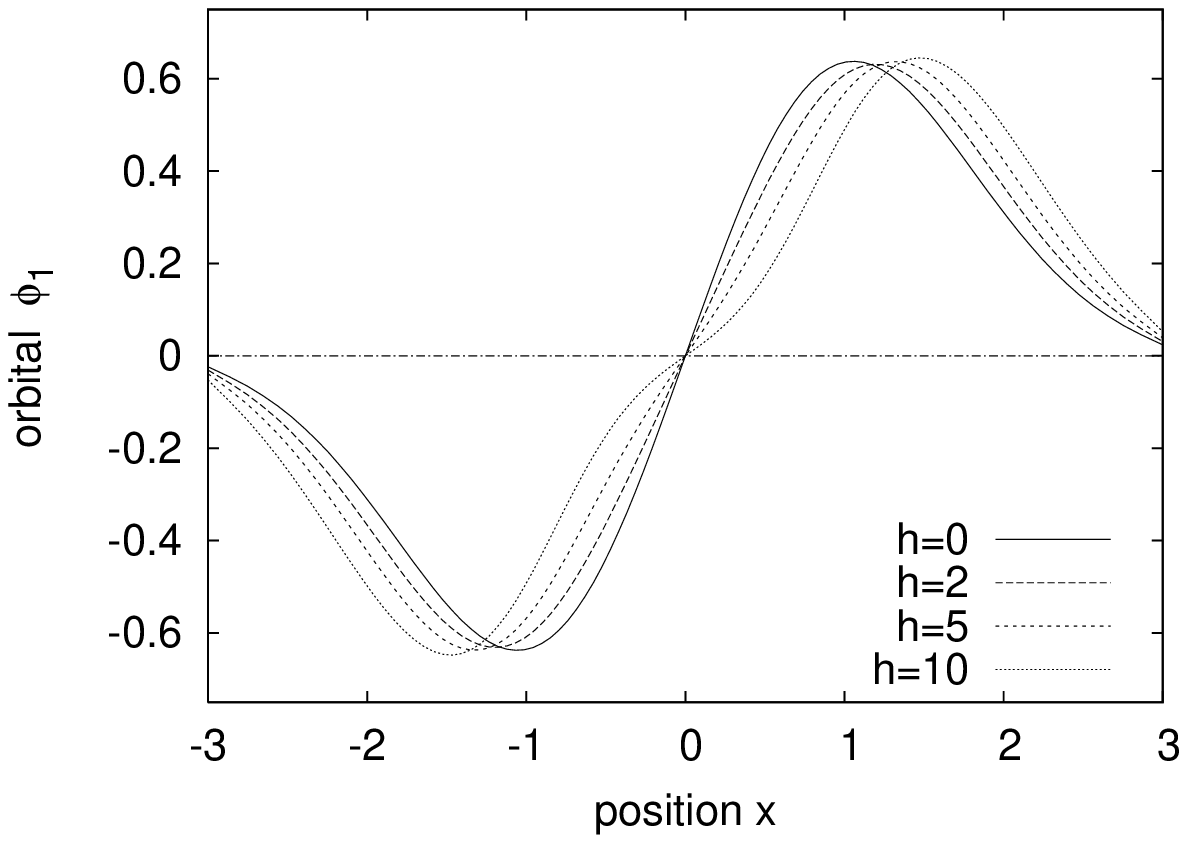}\end{center}

\caption{Evolution of the natural orbitals $\phi_{a}$ as $h\to\infty$ for
the case $N=3$ ($g=0.2$). \emph{Top}: The first symmetric orbital
$\phi_{0}$ is notched at $x=0$. \emph{Bottom}: The antisymmetric
one ($\phi_{1}$) is barely altered. \label{cap:spf_h}}
\end{figure}
First consider the borderline case $g=0$. Then the one-particle occupations
$\boldsymbol{n}$ are conserved for any parameter $h$, so we can
assume number states $|\boldsymbol{n}\rangle$ as eigenstates (glossing
over the fact that, of course, in the instance of degeneracy, the
interaction operator will pick out a unitarily transformed basis diagonal
in $H_{\mathrm{I}}$). Let us start with the harmonic trap ($h=0$),
where the spectrum is arranged in steps of $\Delta_{0}=1$ according
to $E_{\boldsymbol{n}}=\sum_{a}n'_{a}(a+{\scriptstyle \frac{1}{2}})$
and the particles are distributed over the oscillator orbitals $\phi_{a}$.
Now let us switch on a central barrier $h>0$ peaked at $x=0$. Then
each even orbital $a\in2\mathbb{N}$ will be notched at $x=0$, until
its density $\left|\phi_{a}\right|^{2}$ will equal that of the next,
odd orbital $\phi_{a+1}$. Figure \ref{cap:spf_h} gives an illustration
of this by displaying the natural orbitals $\phi_{0/1}$. Along that
line, the energies will evolve continuously from $\epsilon_{a}$ to
$\epsilon_{a+1}=a+{\scriptstyle \frac{3}{2}}$. On the other hand,
granted that the barrier is supported exclusively at $x=0$, the odd
orbitals themselves will remain completely untouched. Hence, in the
limit $h\to\infty$, we would end up with a doubly degenerate \emph{single-particle}
spectrum (or, more realistically, a level gap $\Delta_{h}\ll1$),
which readily translates to a shift of $\Delta E_{\boldsymbol{n}}=\sum_{a\in2\mathbb{N}}n'_{a}\times1=:n'_{\mathrm{even}}$,
depending on how many even orbitals were populated to begin with.
Altogether, as the barrier $h$ is run up, the spectrum $\{ N/2,N/2+1,\dots\}$
at $h=0$ is expected to transform into one with a lowest cluster
of $1+N$ (quasi-)degenerate levels pertaining to $\{|n'_{0},N-n'_{0}\rangle\}$
at energies $E\sim3N/2$, followed by another one at $E\sim3N/2+2$. 

\begin{figure}
\includegraphics[%
  width=6.7cm,
  keepaspectratio]{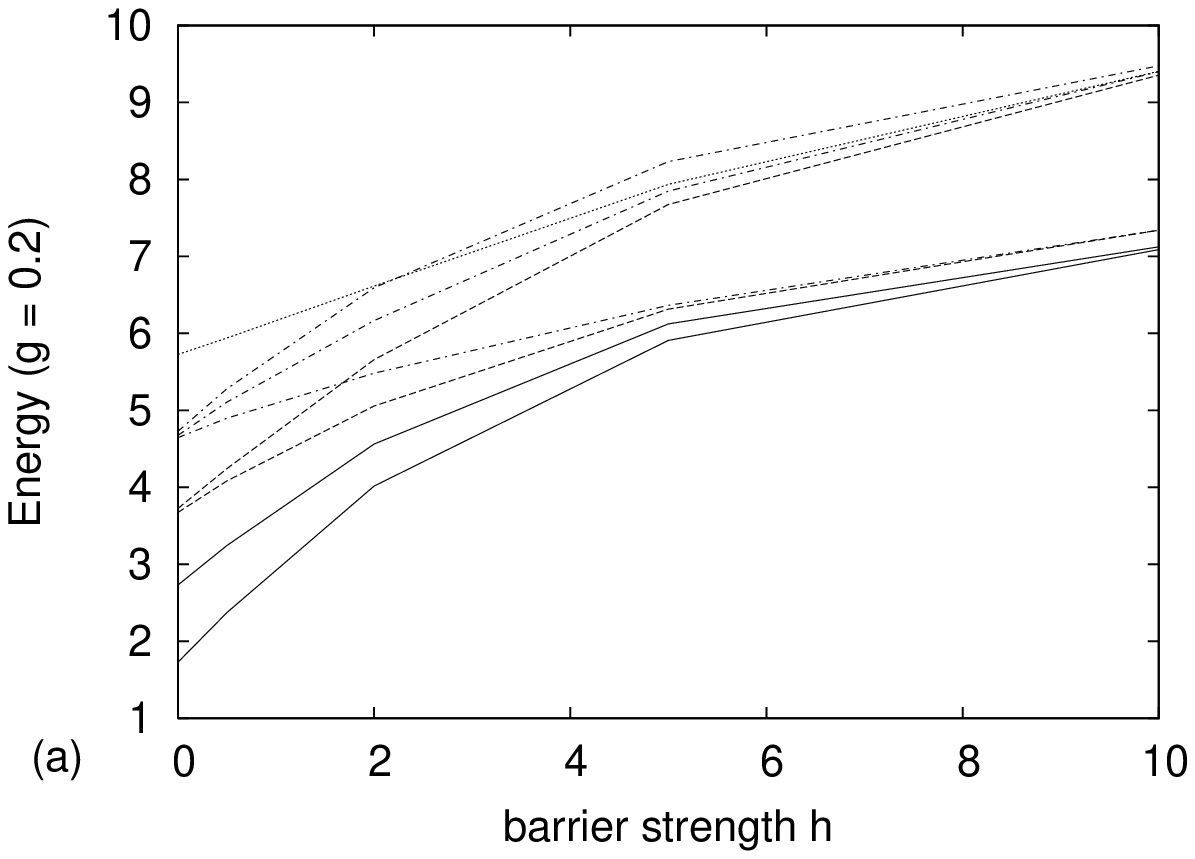}

\includegraphics[%
  width=6.7cm,
  keepaspectratio]{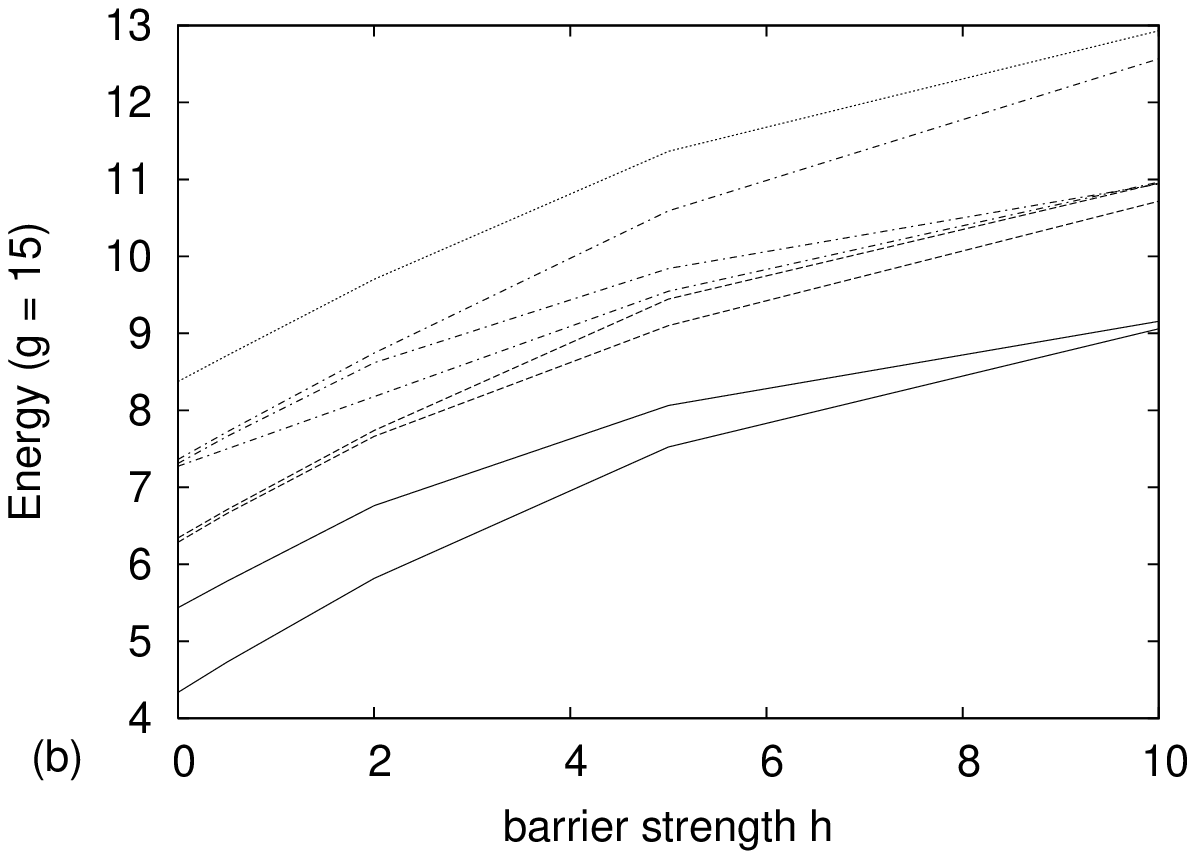}

\caption{Crossover of the lowest energies $E_{m}(h)$ with varying barrier
strength $h$ for $N=3$ bosons at interaction strengths $g=0.2$
(a); $g=15$ (b). The line styles are assigned so as to distinguish
the different level groups at $h=0$. \label{cap:crossover_h}}
\end{figure}

It goes without saying that a realistic reasoning should take into
account the finite barrier width ($w=0.5$), but the above toy model
provides us with a rough picture to understand the crossover computed
for $g=0.2$ in Fig.~\ref{cap:crossover_h}(a). Note that the sketched
metamorphosis inevitably brings about crossings between different
levels as $h\to\infty$ since, for instance, $|0,N\rangle$ is barely
altered while $|n'_{0},0,N-n'_{0}\rangle$ is shifted by about $\Delta E\sim N$. 

The above approach may be readily extended to the fermionization limit.
All we need to do is construct auxiliary fermion states $\{|\boldsymbol{n}\rangle\mid n'_{a}=0,1\}$
and apply the same machinery. However, a look at Fig.~\ref{cap:crossover_h}(b)
($g=15$) makes clear that the rearrangement of the levels is not
as wild as as in the non-interacting case. That is simply because
the {}`fermions' can only occupy a level once; hence at $h\to\infty$
the lowest group is made up of one or two states only (for even/odd
numbers, respectively), followed by a cluster of four levels regardless
of the atom number. You might notice that the second band emerging
as $h\to\infty$ is not perfectly bunched at $E(h=10)\simeq11$, but
really has a runaway at $E(h=10)\simeq10.7$. This can be traced back
to the inclusion of a higher orbital $\phi_{4}$ in the fermionic
state: in such higher regions, the spectrum ceases to be perfectly
doublet-like, foiling our previous considerations.

For intermediate values of $g$, in turn, one cannot use the same
line of argument since the interaction is in the way of a simple single-particle
description, and $\boldsymbol{n}$ are no longer good quantum numbers.
Still, the knowledge of the limiting cases highlighted above gives
a guideline for the crossover. Generally speaking, changing $h$ for
any $g$ will affect the energy via \[
\frac{d}{dh}E=N\mathrm{tr}\left(\rho_{1}\delta_{w}(x)\right)=N\bar{\rho}(0),\]
i.e, the coarse-grained density $\bar{\rho}\equiv\rho*\delta_{w}$
about the center will be reduced so as to minimize the energy costs.
This will determine the fate of each state when changing over from
a single to a double well, thus completing our picture of the lowest
excitations in double-well traps.

\section{Conclusions and outlook}

We have examined the lowest excitations of $N\le5$ bosons in harmonic
and double-well traps, based on the numerically exact Multi-Configuration
Time-Dependent Hartree method. The key aspect has been the spectral
evolution from the weakly to the strongly interacting limit, this
way extending our previous analyses of the ground state \cite{zoellner06a,zoellner06b}
to the lowest excitations. Moreover, we have illuminated the crossover
from a single well to a pronounced double well.

In the case of a purely harmonic trap, the initially equidistant and
degenerate level structure is lifted as interactions are introduced,
which distinguish between different states of the relative motion.
In the \emph{fermionization} limit of ultrastrong repulsion, a harmonic
spectrum is recovered asymptotically. In a double well, the non-interacting
spectrum has a lowest band composed of $N+1$ states formed from the
(anti-)symmetric single-particle orbital, well separated from the
next upper band. Here, the effect of interactions consists in a complex
rearrangement of the levels, dominated by level repulsion in the perturbative
regime. Moreover, some lines virtually adhere to one another as interactions
are switched on, despite their being very different in character.
In the fermionization limit, we end up with a lowest group made up
of the ground state (even atom numbers) plus the first excited state
(odd), followed by a cluster of four levels for any $N$, that washed
out due to the non-doublet nature of the higher-lying orbitals.

In order to get a better understanding, we have also analyzed the
underlying eigenvectors $\Psi_{m}$. At bottom, the same mechanism
responsible for the ground-state fermionization could be identified.
Stronger interactions imprint a two-body \emph{correlation hole,}
signifying a reduced probability of finding two particles at the same
position, and eventually lead to localization. This becomes visible
in the density profiles, which evolve from a Gaussian envelope to
a significantly flatter shape. However, the excited states elude an
intuitive interpretation applicable to the ground state.

Finally, we have cast a light on just how the spectrum is reorganized
when splitting the purely harmonic trap into two fragments. To this
end, we considered the deformation of the single-particle orbitals
as a central barrier is run up. This leads to very different energy
shifts depending on the overall population of even orbitals or, generally,
the average density about the trap's center.\\

With these systematic investigations, we have complemented the extensive
work on the ground state. Numerically delicate as it is, our study
has been limited to the lowest excitations and also to at most five
atoms with an eye toward computing time. On the other hand, we hope
it will also contribute to the understanding of dynamical but also
thermal properties. In this light, an obvious extension would be to
study time-dependent phenomena. Here double-well systems have proven
to be a fruitful model for various phenomena.

\begin{acknowledgments}
Financial support by the Landesstiftung Baden-Württemberg in the framework
of the project {}`Mesoscopics and atom optics of small ensembles
of ultracold atoms' is gratefully acknowledged by P.S. and S.Z. The
authors also thank O.~Alon for valuable discussions.
\end{acknowledgments}
\bibliographystyle{prsty}
\bibliography{/home/sascha/paper/pra/DW/phd,/home/sascha/bib/mctdh}

\begin{thebibliography}{10}

\bibitem{pitaevskii}
L. Pitaevskii and S. Stringari, {\em Bose-Einstein Condensation} (Oxford
  University Press, Oxford, 2003).

\bibitem{dalfovo99}
F. Dalfovo, S. Giorgini, L. Pitaevskii, and S. Stringari, Rev. Mod. Phys. {\bf
  71},  463  (1999).

\bibitem{pethick}
C.~J. Pethick and H. Smith, {\em Bose-Einstein condensation in dilute gases}
  (Cambridge University Press, Cambridge, 2001).

\bibitem{leggett01}
A.~J. Leggett, Rev. Mod. Phys. {\bf 73},  307  (2001).

\bibitem{Olshanii1998a}
M. Olshanii, Phys. Rev. Lett. {\bf 81},  938  (1998).

\bibitem{girardeau60}
M. Girardeau, J. Math. Phys. {\bf 1},  516  (1960).

\bibitem{vaidya79}
H.~G. Vaidya and C.~A. Tracy, Phys. Rev. Lett. {\bf 42},  3  (1979).

\bibitem{minguzzi02}
A. Minguzzi, P. Vignolo, and M.~P. Tosi, Phys. Lett. A {\bf 294},  222  (2002).

\bibitem{girardeau01}
M. Girardeau, E.~M. Wright, and J.~M. Triscari, Phys. Rev. A {\bf 63},  033601
  (2001).

\bibitem{papenbrock03}
T. Papenbrock, Phys. Rev. A {\bf 67},  041601  (2003).

\bibitem{kinoshita04}
T. Kinoshita, T. Wenger, and D.~S. Weiss, Science {\bf 305},  1125  (2004).

\bibitem{paredes04}
B. Paredes {\it et~al.}, Nature {\bf 429},  277  (2004).

\bibitem{petrov00}
D.~S. Petrov, G.~V. Shlyapnikov, and J.~T.~M. Walraven, Phys. Rev. Lett. {\bf
  85},  3745  (2000).

\bibitem{lieb03}
E.~H. Lieb, R. Seiringer, and J. Yngvason, Phys. Rev. Lett. {\bf 91},  150401
  (2003).

\bibitem{sakmann05}
K. Sakmann, A.~I. Streltsov, O.~E. Alon, and L.~S. Cederbaum, Phys. Rev. A {\bf
  72},  033613  (2005).

\bibitem{hao06}
Y. Hao, Y. Zhang, J.~Q. Liang, and S. Chen, Phys. Rev. A {\bf 73},  063617
  (2006).

\bibitem{Busch98}
T. Busch, B.~G. Englert, K. Rzazewski, and M. Wilkens, Found. Phys. {\bf 28},
  549  (1998).

\bibitem{blume02}
D. Blume, Phys. Rev. A {\bf 66},  053613  (2002).

\bibitem{alon05}
O.~E. Alon and L.~S. Cederbaum, Phys. Rev. Lett. {\bf 95},  140402  (2005).

\bibitem{deuretzbacher06}
F. Deuretzbacher, K. Bongs, K. Sengstock, and D. Pfannkuche, cond-mat/0604673
  (2006).

\bibitem{streltsov06}
A.~I. Streltsov, O.~E. Alon, and L.~S. Cederbaum, Phys. Rev. A {\bf 73},
  063626  (2006).

\bibitem{zoellner06a}
S. Z{\"o}llner, H.-D. Meyer, and P. Schmelcher, Phys. Rev. A {\bf 74},  053612
  (2006).

\bibitem{zoellner06b}
S. Z{\"o}llner, H.-D. Meyer, and P. Schmelcher, Phys. Rev. A {\bf 74},  063611
  (2006).

\bibitem{masiello05}
D. Masiello, S.~B. McKagan, and W.~P. Reinhardt, Phys. Rev. A {\bf 72},  063624
   (2005).

\bibitem{masiello06}
D.~J. Masiello and W.~P. Reinhardt, cond-mat/0610609  (2006).

\bibitem{cederbaum04}
L.~S. Cederbaum and A.~I. Streltsov, Phys. Rev. A {\bf 70},  023610  (2004).

\bibitem{andrews97}
M.~R. Andrews {\it et~al.}, Science {\bf 275},  637  (1997).

\bibitem{shin04}
Y. Shin {\it et~al.}, Phys. Rev. Lett. {\bf 92},  050405  (2004).

\bibitem{anker05}
T. Anker {\it et~al.}, Phys. Rev. Lett. {\bf 94},  020403  (2005).

\bibitem{mey03:251}
H.-D. Meyer and G.~A. Worth, Theor.\ Chem.\ Acc. {\bf 109},  251  (2003).

\bibitem{mey98:3011}
H.-D. Meyer,  in {\em {T}he {E}ncyclopedia of {C}omputational {C}hemistry},
  edited by P. v.~R.~Schleyer {\it et~al.} (John Wiley and Sons, Chichester,
  1998), Vol.~5, pp.\ 3011--3018.

\bibitem{bec00:1}
M.~H. Beck, A. J{\"a}ckle, G.~A. Worth, and H.-D. Meyer, Phys.\ Rep. {\bf 324},
   1  (2000).

\bibitem{yukalov05}
V.~I. Yukalov and M.~D. Girardeau, cond-mat/0507409  (2005).

\bibitem{mctdh:package}
G.~A. Worth, M.~H. Beck, A. J{\"a}ckle, and H.-D. Meyer, The {MCTDH} {P}ackage,
  {V}ersion 8.2, (2000). H.-D. Meyer, {V}ersion 8.3 (2002). {S}ee
  http://www.pci.uni-heidelberg.de/tc/usr/mctdh/.

\bibitem{kos86:223}
R. Kosloff and H. Tal-Ezer, Chem.\ Phys.\ Lett. {\bf 127},  223  (1986).

\bibitem{meyer06}
H.-D. Meyer, F.~L. Qu\'{e}r\'{e}, C. L\'{e}onard, and F. Gatti, Chem. Phys.
  {\bf 329},  179  (2006).

\bibitem{Busch03}
T. Busch and G. Huyet, J. Phys. B {\bf 36},  2553  (2003).

\bibitem{kolomeisky00}
E.~B. Kolomeisky, T.~J. Newman, J.~P. Straley, and X. Qi, Phys. Rev. Lett. {\bf
  85},  1146  (2000).

\end{thebibliography}

\end{document}